\documentclass[journal]{IEEEtran}

\usepackage{amsmath,enumitem}
\usepackage{amsfonts}
\usepackage{amssymb}
\usepackage{xcolor}
\usepackage{amsthm}
\usepackage{bm}
\usepackage{mathrsfs}
\usepackage{array}
\usepackage{algorithmicx}
\usepackage{algorithm,algpseudocode}
\usepackage{algpseudocode}
\usepackage{graphicx,hyperref}
\usepackage{subfigure}
\usepackage{booktabs}
\usepackage{tikz}
\usetikzlibrary{arrows,shapes}
\usepackage{pgfplots}
\pgfplotsset{compat=1.9}
\renewcommand{\figurename}{Figure}

\definecolor{orange}{RGB}{255,127,0}
\definecolor{byzantine}{rgb}{0.74, 0.2, 0.64}
\definecolor{airforceblue}{rgb}{0.36, 0.54, 0.66}
\definecolor{cadmiumgreen}{rgb}{0.0, 0.42, 0.24}
\definecolor{cadmiumorange}{rgb}{0.93, 0.53, 0.18}
\definecolor{bazaar}{rgb}{0.6, 0.47, 0.48}
\definecolor{darklavender}{rgb}{0.45, 0.31, 0.59}
\definecolor{darkcerulean}{rgb}{0.03, 0.27, 0.49}
\definecolor{jazzberryjam}{rgb}{0.65, 0.04, 0.37}

\newtheorem{prop}{\textbf{Proposition}}
\newtheorem{lem}{\textbf{Lemma}}
\newtheorem{definition}{\textbf{Definition}}
\newtheorem{remark}{\textbf{Remark}}

\begin{document}

%\doublespacing
\title{DOA Estimation via \\ Optimal Weighted Low-Rank Matrix Completion}

\author{Saeed~Razavikia, Mohammad~Bokaei, Arash~Amini, Stefano~Rini and Carlo Fischione

\thanks{S. Razavikia is with the School of Electrical Engineering and Computer Science,  KTH Royal Institute of Technology, Stockholm, Sweden (e-mail: \href{mailto:sraz@kth.se}{sraz@kth.se}). 

M. Bokaei is with the Department of Electronic Systems, Aalborg University, Aalborg, Denmark,(e-mail:  \href{mailto:mohammadb@es.aau.dk}{mohammadb@es.aau.dk}). 

A. Amini is with the EE Department at Sharif University of 
%an independent researcher,
Technology, Tehran, Iran
(e-mail:   \href{mailto:aamini@sharif.edu}{aamini@sharif.edu}). 
S. Rini is with the Electrical and Computer Engineering Department, National Yang-Ming Chao-Tung University (NYCU), Taiwan 
(e-mail: \href{mailto:stefano.rini@nycu.edu.tw}{stefano.rini@nycu.edu.tw}). 

C. Fischione is with the School of Electrical Engineering and Computer Science and Digital Futures of KTH Royal Institute of Technology, Stockholm, Sweden (e-mail: \href{mailto:carlofi@kth.se}{carlofi@kth.se}). 

S. Razavikia was jointly supported by the Wallenberg AI, Autonomous Systems and Software Program (WASP) and the Ericsson Research Foundation. The EU FLASH project, the SSF SAICOM project, the Digital Futures project DEMOCRITUS, and the Swedish Research Council Project MALEN partially supported this work.

A preliminary version of this work was presented in part at the   European Signal Processing Conference, Belgrade, Serbia, 2022, which appears in this manuscript as reference~\cite{BokaeiDOA2022}.}
}
\maketitle

\begin{abstract}
This paper presents a novel method for estimating the direction of arrival (DOA) for a non-uniform and sparse linear sensor array using the weighted lifted structure low-rank matrix completion.  The proposed method uses a single snapshot sample in which a single array of data is observed. The method is rooted in a weighted lifted-structured low-rank matrix recovery framework. The method involves four key steps: (i) lifting the antenna samples to form a low-rank stature, then  (ii) designing left and right weight matrices to reflect the sample informativeness, (iii) estimating a noise-free uniform array output through completion of the weighted lifted samples, and (iv) obtaining the DOAs from the restored uniform linear array samples. 
We study the complexity of steps (i) to (iii) above, where we analyze the required sample for the array interpolation of step (iii) for DOA estimation. We demonstrate that the proposed choice of weight matrices achieves a near-optimal sample complexity. This complexity aligns with the problem's degree of freedom, equivalent to the number of DOAs adjusted for logarithmic factors. Numerical evaluations show the proposed method's superiority against the non-weighted counterpart and atomic norm minimization-based methods. Notably, our proposed method significantly improves, with approximately a $10$ dB reduction in normalized mean-squared error over the non-weighted method at low-noise conditions.
\end{abstract}


% -----------------------------------------
% ========================================= SR EDITS
% -----------------------------------------

\section{Introduction}
Direction-of-arrival (DOA) estimation retrieves the relative direction of an unknown number of sources from the signal outputs of several receiving antennas that form a sensor array. DOA estimation is useful in several engineering applications, such as radar~\cite{Seidi2022Demixing},  wireless communication~\cite{daei2023blind,razavikia2022blind}, integrated sensing and communications~\cite{daei2024timely}, and noise localization. 

This paper considers a general method for DOA estimation from a sparse, non-uniform linear array of antennas, where the number of antenna pairs with small separations (small multiples of $\lambda/2$) is much smaller than in uniform linear arrays~\cite{guo2018doa}. 
Sparse array configurations, including nested arrays \cite{pal2010nested} and coprime arrays \cite{vaidyanathan2010sparse, pal2011coprime, qin2015generalized}, are favored for their efficiency in direction estimation. Under these settings, we propose a novel method that estimates the DOAs. We show that the proposed method reduces the sample complexity (required array samples for perfect recovery) of estimating the DOAs. Further, simulation results confirm that our method outperforms previous techniques proposed in the literature, such as atomic norm minimization (ANM)~\cite{bhaskar2013atomic,chi2020harnessing}, EMaC method~\cite{chen2014robust,razavikia2019reconstruction,razavikia2019sampling}, double EMaC (DEMaC) method~\cite{yang2021new}, basis pursuit technique~\cite{chen2001atomic}.

\subsection{Literature Review}

Over the past decades, considerable research efforts have been poured into the  DOA estimation problem~\cite{wang2012low,jin2009joint}.  Traditional subspace methods like MUSIC~\cite{schmidt1982signal}, Prony~\cite{prony1795essai}, and ESPRIT~\cite{roy1989esprit}  have received wide attention in the literature.  The aforementioned methods need many snapshots to estimate the covariance matrix of observations accurately and mostly apply to uniform linear arrays (ULA)s~\cite{moghaddamjoo1991application}. Moreover, these subspace methods are likely to fail when coherent, highly correlated, or closely located sources occur~\cite{malioutov2005sparse}. Additionally, subspace methods need to know the number of sources a priori.

To overcome the limitations above, some methods have been proposed based on a covariance matrix sampled signal, such as second-order statistics of eigenvalues~\cite{han2013improved} or information criterion rules~\cite{stoica2004model} that automatically take care of order estimation. While most classical techniques rely on approximating the auto-correlation function of the received signal~\cite{schmidt1982signal, roy1989esprit, kim1996two}, more recent techniques can estimate the source locations based on only one single snapshot (time slot)~\cite{jin2009joint, wang2012low}.

With the emergence of compressed sensing~\cite{donoho2006compressed,candes2006stable}, substantial progress has been achieved in sparse signal recovery.  The sparsity-aware DOA estimation techniques brought the advantage of estimating the directions with only a single snapshot. Interestingly, the latter methods need only an upper bound on the number of sources and not the exact number; the drawback of this method is that the estimated source angles (directions) are constrained to belong to a predefined finite set, which is commonly referred to as the grid~\cite{malioutov2005sparse}. It is common to use sparse recovery techniques in compressed sensing that require discretizing the range of angles (an angular grid) to take advantage of the latter sparsity.  In other words, the range of feasible angles needs to be discretized beforehand; a finely discretized grid is expected to yield more accurate DOAs at the expense of more computational cost~\cite{duarte2013spectral}.

The super-resolution technique of \cite{candes2014towards} provided a grid-less convex formulation for a frequency estimation problem closely linked with the DOA estimation. Instead of a grid resolution requirement, this method requires a minimum angular separation between the sources; nevertheless, it does not restrict the exact direction of the sources. Although this method is more accurate than the standard sparse recovery techniques without a considerable additional computational burden, it is tailored to a uniform linear array. Such a restriction seems too limiting, as uniform arrays are generally inefficient for sparse DOA estimation~\cite{xu2014compressed}. 
Moreover, the concept of virtual arrays~\cite{pal2011coprime} studied in non-uniform arrays shows that certain non-uniform array structures are equivalent to uniform arrays with more sensors in the sparse source recovery. There has been a line of recent research to identify suitable antenna placement within an array to enhance the array's potential in detecting sparse sources. While such arrays can probably improve the accuracy of DOA estimates, the super-resolution technique of~\cite{candes2014towards} becomes no longer applicable, and the grid mismatch error (mismatch between the assumed and the actual grid) resurfaces again~\cite{chi2011sensitivity}. This issue was first addressed in~\cite{tang2013compressed}, where a grid-less sparse recovery method was proposed using ANM for sparse linear arrays (SLA)s.  Recent extensions of ANM techniques mainly focus on non-idealities in array configurations. A regularization-free ANM framework was proposed in \cite{wu2024non,wu2024nonfrequency}, introducing a fast semi-definite programming solver that effectively handles non-uniform array and frequency spacing. Similarly, \cite{zhao2023noncovanm} developed a novel nonconvex ANM method to reduce hardware costs and algorithmic complexity in low-cost passive direction-finding systems. Other advancements include the exploration of lens antenna arrays in \cite{hoang2024low,dong2021atomic}, where ANM was applied to enhance DOA estimation by leveraging the focusing capabilities of lens antennas to mitigate the challenges posed by sparse array configurations. Moreover, \cite{zhang2022doa} extended ANM for coherent sources using coprime arrays, demonstrating improved estimation performance in scenarios with non-uniform spacing. These works highlight the growing interest in adapting ANM to handle practical challenges, including sparse, non-uniform arrays and other real-world non-idealities, further advancing the potential of grid-less DOA estimation.

Inspired by the super-resolution technique, a noise-robust grid-less sparse method was proposed in~\cite{chandrasekaran2012convex} based on ANM. This technique was extended in~\cite{tang2013compressed} to SLAs with missing elements. These methods are fairly accurate and robust under a frequency separation constraint (equivalently, DOA separation), especially in noisy settings. Based on enhanced matrix completion (EMaC), another grid-less method was proposed in~\cite{chen2014robust}, which formulates the DOA problem using Hankel-structured matrix recovery; some of the ideas are borrowed from the matrix pencil algorithm in~\cite{hua1992estimating}. The main point in lifting data to the Hankel matrix structure is the low-rank property when the sources are sparse. Interestingly, compared to ANM methods, low-rank Hankel matrix recovery imposes a less restrictive constraint on the minimum angular separation between the sources. Similar lifted structures are also proposed for recovering frequencies in~\cite{yang2021new,ye2016compressive}.

 The method achieves robust performance in scenarios with non-uniform sampling and reduced computational overhead by incorporating a gradient threshold iteration technique and perturbations to escape saddle points. These recent efforts demonstrate significant progress in adapting ANM-based DOA estimation to practical scenarios with sparse and non-uniform arrays, addressing limitations such as grid mismatch and inefficiencies in traditional uniform linear array designs.

Recently, in \cite{Bokaei2022TwoDOA,BOKAEI2023109253}, we introduced a two-snapshot DOA estimation method for SLAs inspired by the leverage scores concept from the matrix completion framework in \cite{chen2014coherent}. This approach utilizes the first snapshot to compute leverage scores and is then employed to identify the most informative array elements not necessarily included in the observed samples. The measurements from these selected array elements are used in the second snapshot to estimate the DOAs. It is worth noting, however, that in most practical DOA estimation scenarios, the array configuration (e.g., SLA) is typically fixed and cannot be adaptively adjusted. Hence,  the method is restricted by the demand for multi-snapshots (at least two), which makes it inapplicable in some practical settings.

In summary, despite significant progress, the following challenges remain unresolved in the DOA estimation literature:
\begin{itemize}[leftmargin=*]
   % \item \textbf{Multiple snapshots:} Traditional subspace methods, such as MUSIC and ESPRIT, rely on multiple snapshots for accurate covariance matrix estimation. This requirement limits their applicability in single-snapshot or rapidly changing environments.
    
    \item \textbf{Grid and array configuration limitations:} Sparse recovery methods constrained by angular grids suffer from grid mismatch errors, where the estimated source directions are restricted to the grid points rather than the actual continuous directions. On the other hand, grid-less techniques such as ANM address this issue but are primarily designed for ULAs. These techniques often fail to generalize to SLAs.

    \item \textbf{Challenges with closely spaced or coherent sources:} Existing methods struggle when sources are closely spaced or exhibit high coherence. This scenario leads to performance degradation, particularly in subspace methods, which rely on strong separability in the angular domain. 

\end{itemize}

In this work, we attempt to address the above challenges by proposing a weighted lifted-structure low-rank matrix completion framework that leverages adaptive sample weighting to improve recovery performance for sparse linear arrays.

% -----------------------------------------
\subsection{Contributions}
% -----------------------------------------

This paper considers the DOA estimation problem and focuses on the weight optimization problem for the so-called lifted-structure matrix completion method. More specifically, we propose a general method that transforms the SLA samples into ULA samples using lifted structures and low-rank matrix recovery.  The proposed method relies on a sample-adaptive weighting scheme for matrix recovery \cite{BokaeiDOA2022}. The proposed method employs a weighting scheme that adapts to samples to improve matrix recovery performance. In this scheme, the weights reflect the significance of the samples concerning the antenna array configuration. First (i), the array of samples is lifted into a Hankel structured matrix. Then (ii) left and right weight matrices are applied to the lifted samples; the array configuration and rough approximate source angles determine these weights. Successively (iii), the full samples from the uniform linear array are estimated as a low-rank matrix completion problem. Finally, (iv) the DOAs are estimated from the reconstructed uniform linear array samples. In other words, our scheme comprises two main steps: (a) determining the weights and (b) lifting the samples to a structured low-rank matrix over which matrix recovery is performed.  We refer to this method as the  \emph{weighted lifted-structure (WLi) low-rank matrix DOA estimation}.

One of our key contributions is in step (a) above. In this step, we devise a strategy for determining the weights of a weighted nuclear-norm minimization. From a high-level perspective, this strategy determines a cost—in the form of a weighted nuclear norm—for which the observed matrix entries are treated as the highest leverage scores. These leverage scores measure the influence of individual sample values on the estimation of missing entries for matrix completion. 

The resulting optimization problem is solved by an alternating direction method of multipliers (ADMM) algorithm~\cite{boyd2011distributed}. We show that the weighting scheme in (a) reduces the sample complexity of the problem, which corresponds to the number of required array elements for estimating the DOAs. Indeed, the selection of weight matrices results in a scenario where the sample complexity becomes independent of leverage scores and is nearly optimal, except for certain logarithmic factors. The simulation results show that the weighted methods outperform the unweighted techniques, grid-based compressed sensing methods, and ANM methods, especially when the sources are not widely separated. 

Our contribution can be summarized as follows.
\begin{itemize}
    \item \textbf{Weighted nuclear-norm minimization:} We devise a strategy for determining the weights in a weighted nuclear-norm minimization framework, which assigns costs based on the leverage scores of observed matrix entries. These leverage scores quantify the influence of individual samples on matrix completion performance.
    \item \textbf{Low computational complexity solution:} The resultant optimization problem for completing the SLA into ULA is based on nuclear norm minimization.  To reduce the complexity of the singular value decomposition, we propose a singular value decomposition-free approach by deriving an ADMM algorithm for the given optimization problem, ensuring computational efficiency. 
    \item \textbf{Single and Multi-snapshot:}  While our weighting scheme is originally formulated for the single-snapshot case, we also provide a generalization to the multi-snapshot setting by modeling the data via block-Hankel matrices and deriving snapshot-wise diagonal weights, thus extending the algorithm’s utility to multiple measurement vectors.
    \item \textbf{Near optimal sample complexity:} We theoretically demonstrate that the weighting scheme reduces the sample complexity required for DOA estimation. Indeed, the sample complexity becomes independent of leverage scores and is nearly optimal, except for certain logarithmic factors. 
    \item \textbf{Numerical experiments:}  We provide numerical experiments for the proposed weighted methods, outperforming unweighted, grid-based, compressed sensing, and ANM methods. Numerical results reveal that WLi-EMaC's advantage is particularly evident in scenarios with closely spaced sources, significantly reducing the normalized mean squared error.
\end{itemize}

We note that although our scheme is similar to the ANM methods in  \cite{li2015off,yang2015gridless}, it does not need to input the statistics of the sources. In addition, the weighted method applies to single and multi-snapshot scenarios; however, the single-snapshot version is discussed in the following.  Furthermore, unlike traditional methods such as MUSIC and ESPRIT, which require the number of sources as input, the proposed method does not need prior knowledge of the source count. The sparsity-promoting optimization adaptively estimates the underlying DOA structure based on the observed data. After having presented the weighted lifted-structured low-rank matrix completion DOA estimation in some generality, we consider Hankel and double-Hankel structures as special cases (similar to EMaC~\cite{chen2014robust} and DEMac~\cite{yang2021new}, respectively), resulting in our proposed WLi-EMaC and WLi-DEMaC methods.

\subsection{Organization}

The rest of the paper is organized as follows: the signal model and problem formulation for DOA estimation are presented in Section~\ref{Sec:Model}. In Section~\ref{sec:Proposed Approach},  we define our method for DOA estimation in a noisy setup and introduce the proposed weighted matrix completion. We study the weight matrix optimization problem and DOA estimation in Section~\ref{sec:Design the Weight Matrices} and Section~\ref{sec:DOAestiion_ula}, respectively. In Section~\ref{Sec:Simulation}, we give numerical simulations by taking into account the special instances of the proposed method, namely WLi-EMaC and WLi-DEMaC. Finally, Section~\ref{Sec:Conclude} concludes the paper.  

\medskip

%\subsection{Notations}
\noindent
\textbf{Notations:}
We denote scalars, vectors, and matrices by lowercase letters and lower and upper-case boldface letters, respectively. We represent linear operators and their adjoints by calligraphic notations such as $\mathcal{X}$ and $\mathcal{X}^{\dagger}$, where the superscript in the latter stands for the adjoint operator.  We frequently use $\Omega$ as a finite set of integers and denote its cardinality by $|\Omega|$. The transpose and Hermitian of a matrix $\mathbf{X}$ are represented by $\mathbf{X}^{\mathsf{T}}$ and $\mathbf{X}^{\mathsf{H}}$, respectively.  Following the conventional notations, we define $\|\mathbf{X}\|$, $\|\mathbf{X}\|_{\rm F}$ and $\|\mathbf{X}\|_{\rm *}$ as the spectral, Frobenius, and nuclear norms (sum of singular values) of the matrix $\mathbf{X}$, respectively. $\mathbf{e}_i^N$ represents the $i$-th canonical $N$-dimensional basis vector. For an integer $N$, $[N]$ stands for $\{1,2,\dots, N\}$. 
% =================================
\section{System Model}\label{Sec:Model}
% =================================
\subsection{Signal Model}

Let us consider the DOA setting in Fig.~\ref{Fig:DOAschem}: consider $N$ uniformly spaced points on a line, with spacing $s_d$, so that the point locations are $\mathbf{s}=[0,s_d, \ldots, (N-1) s_d]$. Assume that $K$ narrow-band sources are in the surveillance region, respectively located at angles $\{\theta_k\}_{k \in [K]}$, $\theta_k \in [0,2\pi)$.  For each position in $\mathbf{s}$, let $a_n(\theta_k)$ represent the phase shift of the signal from the source $k$ for the $n^{\rm th}$ position. Assuming a narrow-band reflection signal model with wavelength $\lambda$ from the source, the term $a_n(\theta_k)$ is subject to a constant phase shift and thus can be expressed as 
%-------------
\begin{align}
\label{eq:atom}
a_n(\theta_k) = \exp \{{-{\rm j}\tfrac{2\pi}{\lambda}ns_d\sin (\theta_k)}\}.
\end{align}
%------------
Accordingly, the overall signal received at the $n^{\rm th}$ position is obtained as 
%------------
\begin{subequations}
    \label{eq:measurement all}
    \begin{align}
        \label{eq:measured_Discrete}
         y_n &= \sum\nolimits_{k \in [K]}  b_{k}\,  a_n(\theta_{k})  + e_n
          \\
          & =\sum\nolimits_{k \in [K]} b_{k}{\rm e}^{-{\rm j}2\pi\tau_{k}n } + e_n. 
           \label{eq:measurement}
    \end{align}
\end{subequations}
%-------------------
where the coefficients $\{b_{k}\}_{k \in [K]} \in \mathbb{C}$ model the effect of the received power from the sources and their relative phase values concerning a reference. Also, $\{e_{n}\}_{n \in [N]}$ present noise values whose amplitudes are upper-bounded as $|e_n|<\eta$ with high probability. 
The expression in \eqref{eq:measurement} is a commonly-used reformulation of \eqref{eq:measured_Discrete} by defining $\tau_{k} = \frac{s_d}{\lambda}\sin (\theta_{k})$, 
in \eqref{eq:atom}. Note that in \eqref{eq:measured_Discrete}, we have assumed that the narrow-band scattered signals from all the sources have the same wavelength. For the configuration of antenna elements in the array, we consider two scenarios: (i) the uniform linear array (ULA) case in which an antenna is placed in each position of $\mathbf{s}$, and (ii) the non-uniform array (SLA) in which antennas are placed in $M$ out of the $N$ positions in $\mathbf{s}$.

The DOA literature shows that generally speaking, ULAs are simpler to study than other array configurations due to their inherent symmetry. Therefore, we consider a non-uniform array with $M$ elements as an $N$-element ULA with missing antenna arrays and try to recover the missing samples. Let $\Omega \subseteq [N]$ with $|\Omega|=M$ be associated with the existing elements of the ULA in the non-uniform array. A general method to the reconstruction of the samples in the positions $[N] \setminus \Omega$ is through an orthogonal projection $\mathcal{P}_{\Omega}:\mathbb{C}^{N}\to\mathbb{C}^{M}$. More precisely, let us  define $\mathcal{P}_{\Omega}$ as the orthogonal projection that  keeps only the $M$ elements indexed within $\Omega$  of a vector  $\mathbf{y}= [y_1,\ldots,y_N]^{\mathsf{T}}$:  
\begin{align}
\label{eq:Projection}
\mathbf{y}_{\Omega} =\mathcal{P}_{\Omega}(\mathbf{y}).
\end{align} 
The orthogonal projection $\mathcal{P}_{\Omega}$ is determined by the sampling strategy. In other words, \eqref{eq:Projection} describes how the noisy samples at the non-uniform array $\mathbf{y}_{\Omega}$ are obtained as a projection of the vector of ULA noisy samples $\mathbf{y}$. Note that the energy of the noise in $\mathbf{y}_{\Omega}$ is upper-bounded by $\sqrt{|\Omega|}\eta = \sqrt{M}\, \eta$ with high probability.

% =====================
\input{figs/Fig_DOAschem}
% =====================

\subsection{DOA Estimation}

In this section, we introduce the proposed method. For clarity of exposition, we present our algorithm for the noiseless case, i.e., \eqref{eq:measurement all} when the error terms $\{e_n\}$ are absent.
In the noiseless scenario, the source estimation problem in DOA is equivalent to finding the pairs $(\tau_{k}, b_{k})$ for $k \in [K]$, by using the sample vector $\mathbf{y}_{\Omega}$. Although $k$ belongs to a discrete set, the values of $\tau_{k}$ and $b_{k}$ vary continuously. One possible method is discretizing $\tau_{k}$ to convert the problem of DOA estimation into a finite-dimensional form. A direct consequence of the discretization step is that the estimated source angles are forced to lie on a grid. Therefore, the estimated angles are inherently imprecise due to the grid mismatch. A continuous-domain modeling suitable for the DOA problem is proposed in \cite{candes2014towards} as follows.  
%--------------
\begin{align}\label{eq:signal}
x(t) &= \sum\nolimits_{k \in [K]}b_{k}\delta(t -\tau_k),
\end{align}
%------------
where $\delta(\cdot)$ is the Dirac's delta distribution, and $(\tau_k,b_k)$ are the unknown parameters to be estimated from the sample vector $\mathbf{y}_{\Omega}$. Then, the noiseless samples in \eqref{eq:measured_Discrete} can be obtained as $n$-th Fourier samples of the signal $x(t)$, i.e.,  
%------------
\begin{align}
    y_n = \mathcal{F}(x(t))|_{t=n} = \sum_{k=1}^Kb_k{\rm e}^{-{\rm j}2\pi \tau_k n}, \quad n\in [N], 
\end{align}
%------------
where $\mathcal{F}$ denotes the Fourier transform. For the setting in \eqref{eq:signal}, instead of estimating the set $\{(\tau_{k}\,,\, b_{k})\}_{k \in [K]}$, one can consider recovering the function $x(\theta)$ via
%--------------
\begin{align}
\label{convex-TV}
\widehat{x}  = \underset{x}{\rm argmin} ~\| x \|_{\mathsf{TV}}  \quad ~~~{\rm s.t.}~ \|\mathcal{A}(x) - \mathbf{y}\|_{2}^2\leq \sqrt{M} \eta, 
\end{align}
%-------------
where $\|.\|_{\mathsf{TV}}$ stands for the total variation (TV) norm, and $\mathcal{A}$ is the linear operator that maps $x(\cdot)$ into the samples $\mathbf{y}$. As the TV norm promotes sparsity, the minimizer of \eqref{convex-TV} is expected to comprise a few delta distributions. Using the dual formulation, a computationally feasible algorithm for finding the minimizer of \eqref{convex-TV} is proposed in \cite{candes2014towards}. 
This method is applicable when $|\Omega|=N$, i.e., when we have access to the samples $\mathbf{y}$ associated with a ULA. This implies that an efficiency solution to \eqref{convex-TV} can be determined by first (i) estimating the missing elements of the SLA, followed by (ii) applying the minimization method of \cite{candes2014towards}.

% ====================
\input{figs/Fig_Procedure}
% ====================

Having defined the system model and the key challenges in DOA estimation, we present our proposed method, which leverages weighted low-rank matrix completion for improved array interpolation,  in the next section.

\section{Proposed Method and Formulation}\label{sec:Proposed Approach}

Our proposed approach consists of two main components: (i)~interpolating missing array elements using weighted low-rank matrix completion and (ii)~estimating the DOAs from the reconstructed uniform array. The main idea is to convert the SLA sample vector ${\mathbf{y}}_{\Omega}$ into an estimated ULA vector $\hat{\mathbf{y}}$, as we explain in the next section. Once the estimated ULA vector is achieved, one can apply the DOA estimation of~\cite{candes2014towards} to such an estimated vector.  Thus, this paper's major novelty and originality are the proposal of the ULA vector estimation, a challenging and complex task using weighted matrix completion. The schematic of the proposed DOA estimation is depicted in Figure~\ref{fig:Method}.

With the goal of estimating the ULA vector $\hat{y}$ from the SLA vector $y_{\Omega}$, first (i) we lift the vector of observations $y_{\Omega}$ into a structured matrix with repeated elements, then (ii) we recover the missing array samples of the ULA vector by minimizing the rank of such structured matrix. The literature considers various structured matrix choices: Hankel, double-Hankel, wrap-around Hankel, Hankel-block-Hankel, Toeplitz, and multi-level Toeplitz. To establish a cohesive framework containing all such choices, we introduce definitions relevant to lifting operators and matrix completion in the following subsections.

\subsection{Lifting Operator}\label{sec:Lifting operator}

The \emph{lifting operator} transforms a vector into a matrix to gain degrees of freedom. To properly define the class of lifting operators considered in this paper, we initially introduce the concept of the \emph{lifting basis}.

\begin{definition}
\label{Def:LiftingBasis}
We call $\{\mathbf{A}_n\}_{n \in [N]} \subseteq \mathbb{C}^{d_1 \times d_2}$ a \emph{lifting basis} if 
\begin{enumerate}
    \item for all $n \in [N]$, we have that $\|\mathbf{A}_n\|_{\rm F}=1$,
    
    \item all the non-zero elements of $\mathbf{A}_n$ are positive, real and equal, and
    
    \item  $\mathbf{A}_n$s are orthogonal:
    \begin{align}
          {\rm tr}\big(\mathbf{A}_{n_1}^\mathsf{T} \,\mathbf{A}_{n_2}\big) = \delta[n_1-n_2],
    \end{align}
     \item and each column of  $\mathbf{A}_n$ (for $n \in [N]$) has at most one nonzero element, i.e.,
    \begin{align}
         \sum_{j\in[d_2]}\bigg(\sum_{i\in [d_1]}[\mathbf{A}_n]_{i,j}\bigg)^2 =1.
     \end{align}
\end{enumerate}
\end{definition}

For a lifting basis $\{\mathbf{A}_n\}_{n \in [N]}$, Definition \ref{Def:LiftingBasis} implies that the non-zero elements in $\mathbf{A}_n$ are all equal to ${1}/{\sqrt{\|\mathbf{A}_n\|_0}}$. This further shows that
%------------
\begin{align}
\|\mathbf{A}_n\| \leq \|\mathbf{A}_n\|_{\rm F} = 1.
\end{align}
%------------

\begin{definition}
\label{Def:Lifting}
Let $\{\mathbf{A}_n\}_{n \in [N]} \subseteq \mathbb{C}^{d_1 \times d_2}$ be a lifting basis according to Definition \ref{Def:LiftingBasis}. The linear mapping $\mathscr{H}: \mathbb{C}^{N} \mapsto \mathbb{C}^{d_1 \times d_2}$ defined by
%------------
\begin{align}
\label{eq:DefLif}
\mathscr{H}(\mathbf{x})  = 
%\sum_{k=1}^n
\sum_{n\in [N]}
a_n{\rm tr}\big(\mathbf{e}^{N_n\mathsf{T}}\mathbf{x}\big) \mathbf{A}_{n},
\end{align}
%------------
is called a \emph{lifting operator}, where $\{a_n\}_{n \in [N]}\subseteq \mathbb{C}$ are constants. We can  check that
$\mathscr{H}^{\dagger} :\mathbb{C}^{d_1\times d_2}\mapsto \mathbb{C}^{N}$ with 
%------------
\begin{align*}
   \mathbf{M}\in \mathbb{C}^{d_1\times d_2}:~~~  \mathscr{H}^{\dagger}(\mathbf{M}) = \sum_{\substack{n\in [N], a_n\neq 0}} \tfrac{1}{a_n} {\rm tr} \big(\mathbf{A}_n^{\mathsf T} \mathbf{M}\big)  \mathbf{e}^N_n,
\end{align*}
%------------
defines the orthogonal back projection from $\mathbb{C}^{d_1\times d_2}$ into $\mathbb{C}^N$.
\end{definition}	

By tuning the lifting basis, one can achieve various matrix structures in the output of the lifting operator. In our simulations-- see Section~\ref{Sec:Simulation}-- we examine the Hankel and double-Hankel lifting operators, defined as
%------------
\begin{equation}
\label{eq:DefHank}
{\mathsf H}^{\rm  hnkl}_d (\mathbf{x})=\mathscr{H}(\mathbf{x}) := \left[ \begin{matrix}
x_{1} & x_{2}  & \dots& x_{N-d+1} \\
x_{2} & x_{3}  & \dots& x_{N-d+2} \\
\vdots & \vdots & \ddots&  \vdots\\
x_{d} & x_{d+1}  & \dots & x_{N} 
\end{matrix}  \right].
\end{equation}
%------------
and double-Hankel, defined as ${\mathsf H}^{2 \rm  hnkl}_d = \left[  {\mathsf H}^{\rm hnkl}_d \ {\mathsf H}^{\rm hnkl}_d \right]$.

\subsection{Matrix Completion}

Once the array observation $\mathbf{y}_{\Omega}$ and the missing antenna samples of $\mathbf{y}$ have been lifted, the missing array samples can be recovered by minimizing the rank of the lifted matrix.  This matrix completion method can be mathematically formulated as follows:
(i) for the noiseless case, 
%----------------
\begin{equation}
\label{convex-opt}
\begin{aligned}
\widehat{\mathbf{y}} = \underset{ \mathbf{g} \in \mathbb{C}^{N}}{\rm argmin} \quad \| \mathscr{H}({\mathbf{g}}) \|_{*}, \quad 
{\rm s.t.} \quad  \mathcal{P}_{\Omega}( \mathbf{g})  = \mathbf{y}_{\Omega},
\end{aligned}
\end{equation}
and (ii) for the noisy case,
%---------------
\begin{equation}
\label{convex-noisy-opt}
\begin{aligned}
\widehat{\tilde{\mathbf{y}}} =\underset{ \mathbf{g} \in \mathbb{C}^{N}}{\rm argmin} \quad  \| \mathscr{H}({\mathbf{g}}) \|_{*},~ 
{\rm s.t.} ~ \|\mathcal{P}_{\Omega}( \mathbf{g})- \mathbf{y}_{\Omega}\|_2 \leq \sqrt{M}\eta,
\end{aligned}
\end{equation}
%---------------
where $\eta > 0$ was previously introduced as an upper bound for noise amplitudes -- with high probability. 

\begin{remark}
Note that, for the system model in Section \ref{Sec:Model}, the position of the array samples is fixed, which means adaptive sampling is impossible. 
\end{remark}

\section{Converting SLA to ULA}

In this section, we are now ready to propose a method to convert the SLA into ULA samples. Since the position of array samples is fixed, and we do not have the freedom to adjust the position of the array samples, the standard unweighted nuclear norm minimization in \eqref{convex-noisy-opt} and \eqref{convex-opt} are inefficient~\cite{chen2015completing}. Therefore, instead of considering the original setting in \eqref{convex-noisy-opt}, we consider a variation that introduces two weight matrices-- $\mathbf{W}_{L}$ and $\mathbf{W}_{R}$--  to modify the recovery cost as $\mathbf{W}_{L}\mathscr{H}({\mathbf{g}})\mathbf{W}_{R}^{{\mathsf{H}}}$ consistent with the available non-uniform samples. A more rigorous definition of these weight matrices will be provided later. 

Accordingly, the weighted lifted-structured low-rank matrix recovery of problem \eqref{convex-noisy-opt} is redefined by incorporating left and right weight matrices into the nuclear norm cost function as
% ===============
\begin{align}
\nonumber
\widehat{\mathbf{y}} = & \underset{ \mathbf{g} \in \mathbb{C}^{N}}{\rm argmin}~\| \mathbf{W}_{L}\mathscr{H}({\mathbf{g}})\mathbf{W}_{R}^{{\mathsf{H}} }\|_{*},~~ \\ &{\rm s.t.}~~\| \mathcal{P}_{\Omega}( \mathbf{g} ) -\mathbf{y}_{\Omega} \|_{\rm F} \leq { \sqrt{M}}\eta, \label{weighted-convex-noisy}
\end{align}
% ===============
where $\mathbf{g}$ stands for the variable for full array elements,  $\mathbf{W}_L \in \mathbb{C}^{d\times d}$ and $\mathbf{W}_R \in \mathbb{C}^{(N-d+1)\times (N-d+1)}$ are weight matrices, and $\eta > 0$ was previously introduced as an upper-bound for noise amplitudes (with high probability). The optimization in~\eqref{weighted-convex-noisy} is convex and can be solved using CVX~\cite{cvx}. However, the main issue now becomes the determination of the weight matrices.

In the next subsection, we propose a method for designing the weight matrices, $\mathbf{W}_R$ and $\mathbf{W}_L$, to reduce the sample complexity and, accordingly, to improve the performance of the resulting DOA estimation. 

\begin{remark}
    Note that the weighted low-rank matrix completion in \eqref{weighted-convex-noisy} does not require prior knowledge of the number of sources. Instead, the optimization promotes sparsity in the recovered uniform array samples through nuclear-norm minimization, inherently adapting to the number of significant sources in the data.
\end{remark}

\subsection{Design of the Weight Matrices}\label{sec:Design the Weight Matrices}

Here, our objective is to formulate weight matrices to reduce sample complexity. This entails a definition of sample complexity in terms of leverage scores. Given a specific set of observed samples, denoted as $\Omega$, we propose an optimization framework to decrease the leverage scores and, consequently, the sample complexity. The optimization problem is subsequently recast to focus on weight matrices, enabling us to determine weights directly associated with minimizing leverage scores or sample complexity.

In \cite{BOKAEI2023109253,chen2015completing}, it has been established that the complexity associated with resolving the optimization problem in ~\eqref{weighted-convex-noisy} is directly correlated with the average leverage scores, which are defined as follows.

%----------------
\begin{definition}
\label{LEV_SCORS_M}
For given weight matrices $\mathbf{W}_{L}, \mathbf{W}_{R}$ and an arbitrary vector 
$\mathbf{x} \in \mathbb{C}^{N}$,  assume  that the rank of $\mathbf{W}_{L} \mathscr{H}(\mathbf{x}) \mathbf{W}_{R}^{\mathsf{H}}$ is  $\tilde{K}$. Further, let ${\mathbf{U}}{\boldsymbol{\Sigma}}{\mathbf{V}}^{\mathsf{H}}$ be the reduced  SVD  thereof. 
For each $n \in[N]$, \cite{BOKAEI2023109253,chen2015completing} define the weighted leverage scores  $\mu_{n}$ as
%------------
\begin{align}
\label{eq:mutildDef}
\mu_{n} :=\frac{N}{\tilde{
K}} \max\{\|  \mathcal{P}_{U} (\mathbf{A}_{n})\|_{\rm F}^2,\| \mathcal{P}_{V}(\mathbf{A}_{n}) \|_{\rm F}^2   \}, ~~n \in[N],
\end{align}
%------------
where $\mathcal{P}_{U}(\mathbf{Y})$ and $\mathcal{P}_{V}(\mathbf{Y})$ for arbitrary $\mathbf{Y} \in \mathbb{C}^{d \times (N-d+1)}$ are defines as:
%----------------
\begin{subequations}
    \label{eq:PuPV}
    \begin{align}
        \mathcal{P}_{U}(\mathbf{Y}) &= \mathbf{W}_{L}^{\mathsf{H}} \mathbf{U} \left( \mathbf{U}^{\mathsf{H}} \mathbf{W}_{L} \mathbf{W}_{L}^{\mathsf{H} } \mathbf{U} \right)^{-1} \mathbf{U}^{\mathsf{H}} \mathbf{W}_{L}\mathbf{Y},\\ 
\mathcal{P}_{V}(\mathbf{Y}) &=  \mathbf{Y}\mathbf{W}_{R}^{\mathsf{H}} \mathbf{V} \left( \mathbf{V}^{\mathsf{H}} \mathbf{W}_{R} \mathbf{W}_{R}^{\mathsf{H} } \mathbf{V} \right)^{-1} \mathbf{V}^{\mathsf{H}} \mathbf{W}_{R}.
    \end{align}
\end{subequations}
\end{definition}
%----------------
The rationale of Definition~\ref{LEV_SCORS_M} is that it allows us to characterize the sample complexity. It is shown in \cite{BOKAEI2023109253} that the sample complexity of $\mathcal{O}(\sum_{n}\mu_{n}K^2\log^3{(N)})$   is required to guarantee the perfect recovery of DOAs for the noiseless scenario using optimization \eqref{weighted-convex-noisy} when sampling each element of the ULA proportional to its corresponding score. Note that this complexity is for the case in which the sampling strategy is random, while, in the current setting, the sampling strategy is deterministic.

Accordingly, to reduce the upper bound on the sample complexity, we need to design the weight matrices to minimize the overall leverage scores. In the deterministic sampling strategy, the strategy is to maximize the likelihood of observed samples in $\Omega$. As a result, the actual $\Omega$ can be considered a realization of a random sampling set with elementwise probabilities $\{p_n\}_{n=1}^N$. Accordingly, our method considers the given set of array indexes-- $\Omega$-- as a realization of a random sampling. It can be verified that the likelihood of observing $\Omega$ in such a random setting is given by $\Big(\prod_{n\in \Omega}p_n \Big)\Big(\prod_{n\not\in \Omega}(1-p_n)\Big)$, where  $p_n$ is the probability of observing the sample $n$ of the array.  Next, we would like to maximize this likelihood by tuning the set $\{p_n\}_{n \in [N]}$. 
Note that the trivial solution $p_n=1,~ n\in \Omega$ and $p_n=0,~n\not\in \Omega$ is not consistent with given samples in $\Omega$. By taking the lower-bound of $\{p_n\}$ in  \cite[Theorem 1]{BOKAEI2023109253} into account, we see that the highest likelihood happens when
%------------
\begin{align}\label{eq:p_i}
p_n = \left\{\begin{array}{ll}
1, & n\in\Omega,\\
\min \{ 1 ~,~ c \,  \frac{\mu_{n}\tilde{K}^2}{N}   \log^3{(N)} \}, & n\not\in \Omega.
\end{array} \right.
\end{align}
%------------
For $n\not\in \Omega$, we expect $c \, \frac{\mu_{n}\tilde{K}^2}{N}   \log^3{(N)}$ to be small; otherwise, the overall likelihood shall not be enough to consider $\Omega$ a typical outcome of the random sampling. Therefore, we assume 
$c \, \frac{\mu_{n}\tilde{K}^2}{N}   \log^3{(N)} < 1$ for tuning the weights; nevertheless, we need to recheck this assumption after tuning the weight matrices. Further, if $p_n$ is small enough for $n\not \in \Omega$, we can use the approximation $\sum_{n\not\in\Omega} \log(1-p_n) \approx -\sum_{n\not\in\Omega} p_n$. Our strategy to determine weight matrices $\mathbf{W}_L,\mathbf{W}_R$ is to maximize the likelihood of observing $\Omega$ in the suitable random sampling strategy given in \eqref{eq:Projection}. 
Note that the parameters $\mu_{n}$ involved in \eqref{eq:p_i} depend on the weight matrices. In summary, our strategy to determine $\mathbf{W}_L,\mathbf{W}_R$ is
%------------
\begin{align}
\hspace{-2pt}{\mathbf{W}}_{L}, {\mathbf{W}}_{R}  = \hspace{-1pt}
\underset{\substack{ \mathbf{W}_{L} \in \mathbb{C}^{d\times d},\\ \mathbf{W}_{R} \in  \mathbb{C}^{d' \times d'}} }{\rm argmax} 
\hspace{-1pt}-\sum_{n \not\in \Omega} p_{n} 
\equiv \hspace{-1pt} \underset{\substack{ \mathbf{W}_{L} \in \mathbb{C}^{d\times d},\\ \mathbf{W}_{R} \in  \mathbb{C}^{d' \times d'}} }{\rm argmin} 
\sum_{n \not\in \Omega} \mu_{n},
\label{weight_cal_1}
\end{align}
%------------
where $d'=N-d+1$. 
We recall that $ \mu_{n}$s defined in \eqref{eq:mutildDef} depend on the matrices $\mathbf{U}$ and $\mathbf{V}$, which are in turn, determined by the optimal solution $\hat{\mathbf{y}}$ of optimization in \eqref{convex-opt}. To simplify \eqref{weight_cal_1}, we restrict the weight matrices to be diagonal and real-valued as follows.
%------------
\begin{align}
\label{weight_cal_2}
\{w_{L,i}\}_{i\in [d]},~\{w_{R,i}\}_{i\in [d'] } =  \underset{w_{L,i}, w_{R,i} \in \mathbb{R}_+ }{\rm argmin} 
\sum_{n \not\in \Omega} \mu_{n}, 
\end{align}
%------------
where $w_{L,i}$ and $w_{R,i}$ denote the \emph{square} of the $i$-th diagonal elements of  $\mathbf{W}_L$ and $\mathbf{W}_R$, respectively.  Note that the cost function in \eqref{weight_cal_2} does not have a unique minimizer\footnote{Notice that the set 
$\{\mu_{n}\}$ remains unchanged by constant scaling of the weight matrices, as the projection matrices in \eqref{eq:PuPV} are scale-invariant with respect to weight matrices. 
}. To provide uniqueness and a closed-form solution, we set the sum of weights to be a constant as follows
%------------
\begin{align}
\nonumber
& \{w_{L,i}^*\}_{i\in [d]},~\{w_{R,i}^*\}_{i \in [d']}    
= \underset{w_{L,i}, w_{R,i} \in \mathbb{R}_+ }{\rm argmin} 
\sum_{n \not\in \Omega} \mu_{n}, \nonumber \\ 
& \qquad \qquad \qquad  \text{s.t.}~ \sum_{i\in [d]} w_{L,i} = 1, ~\sum_{i \in [d'] }w_{R,i} = 1.\label{eq:weight_cal_3}
\end{align}
%------------
Next, we use the following proposition to minimize a surrogate function of the leverage scores for the diagonal and real-valued weight matrices.
\begin{prop} [\cite{BOKAEI2023109253} - Corollary 2]
\label{col:diag_weight}
In Definition \ref{LEV_SCORS_M}, if $\mathbf{W}_L \in \mathbb{R}_+^{d\times d}$ and $\mathbf{W}_R \in \mathbb{R}_+^{(d')\times (d')}$ are restricted to be non-negative-valued diagonal matrices,  i.e., 
%------------
\begin{subequations}
    \begin{align}
     \label{eq:wighetddiag}
\mathbf{W}_L & := {\rm diag}\big(\sqrt{w_{L,1}},\dots \sqrt{w_{L,d}}\big), \\
\mathbf{W}_R & := {\rm diag}\big( \sqrt{w_{R,1}} ,\dots, \sqrt{w_{R,d'}}\big).
\end{align}
\end{subequations}
%------------
Then, the leverage scores are bounded by
%------------
\begin{align}
\label{diag_modified_lev_scores3}
\frac{\mu_{n}\tilde{K}}{N} \leq  \max \Bigg\{\frac{\left\| \mathbf{W}_{L} \mathbf{A}_n \right\|_{\rm F}^2 }{\sum_{k=1}^{ \lfloor \frac{N}{\beta \tilde{K}} \rfloor}   w_{L,i_k} } , \frac{\left\| \mathbf{A}_n \mathbf{W}_{R}^{\mathsf{T} }  \right\|_{\rm F}^2 }{ 
\sum_{k=1}^{ \lfloor \frac{N}{\beta \tilde{K}} \rfloor  } w_{R,j_k} }\Bigg\},
\end{align}
%------------
where  {$w_{L,i_1}  \leq \ldots \leq w_{L,{i_{d}}} $ and $w_{R,j_{1}}  \leq \ldots \leq w_{R,j_{d'}}$} are sorted squared diagonal elements of $\mathbf{W}_{L}$ and $\mathbf{W}_{R}$, respectively and 
%--------------
\begin{align}
  \beta = \tfrac{N}{\tilde{K}}\max \{ {1}/{\| \mathbf{U}^{\mathsf{H}} \|^2},  {1}/{\| \mathbf{V}^{\mathsf{H}} \|^2}\}. 
\end{align}
%--------------
\end{prop}

Proposition \ref{col:diag_weight} enables us to use the upper-bound given in \eqref{diag_modified_lev_scores3}, i.e., the objective function in \eqref{eq:weight_cal_3} can be upper-bounded by the following surrogate function. 
%---------------
\begin{align}
  \sum_{n \not\in \Omega} \mu_n\leq 
 \sum_{n \not\in \Omega} \max \big( {  \tfrac{\|\mathbf{W}_L\mathbf{A}_n \|_{\rm F}^2}{\upsilon_{L,N}^{\beta}} } , {  \tfrac{\|\mathbf{W}_R\mathbf{A}_n^{\mathsf{T}} \|_{\rm F}^2}{\upsilon_{R,N}^{\beta}} }\big),    
\end{align}
%---------------
where $\upsilon_{L,N}^{\beta}:=\sum_{i \in \left[ \lfloor N/(\beta \tilde{K})\rfloor \right]} w_{L, i}$ and $\upsilon_{R,N}^{\beta}:= \sum_{j \in \left[ \lfloor N/(\beta \tilde{K})\rfloor \right] } w_{R,j}$. Next, we further bound the denominator of $\tilde{\mu}_n$s  as $\sum_{i \in \left[ \lfloor N/(\beta \tilde{K})\rfloor \right]} w_{L, i} \leq \|\mathbf{W}_{L}\|_{\rm F}^2$ and $ \sum_{j \in \left[ \lfloor N/(\beta \tilde{K})\rfloor \right] } w_{R,j} \leq \|\mathbf{W}_{R}\|_{\rm F}^2$, for the sake of simplicity. Accordingly, we obtain the following  
%---------------
\begin{align}
\label{eq:sum_mu_upper}
  \sum_{n \not\in \Omega} \mu_n\leq 
 \sum_{n \not\in \Omega} \max \big( {  \tfrac{\|\mathbf{W}_L\mathbf{A}_n \|_{\rm F}^2}{\|\mathbf{W}_{L}\|_{\rm F}^2} } , {  \tfrac{\|\mathbf{W}_R\mathbf{A}_n^{\mathsf{T}} \|_{\rm F}^2}{\|\mathbf{W}_{R}\|_{\rm F}^2} }\big),    
\end{align}
%---------------
Because weight matrices are diagonal, the constraint in \eqref{weight_cal_1} simply implies that the denominators of the surrogate function in \eqref{eq:sum_mu_upper} become one in the optimization problem. Hence,  we replace the value of the elements in $\{\mu_{n}\}$ in \eqref{weight_cal_1} with their lower-bounds \eqref{diag_modified_lev_scores3}, optimization problem \eqref{eq:weight_cal_3} transforms into
%------------
\begin{align}
\label{eq:weightedUpper}
 & \mathbf{W}_L^*,~\mathbf{W}_R^*  \nonumber \\
& \quad = \underset{\substack{ \mathbf{W}_L, \mathbf{W}_R \in \mathbb{R}_{+}} }{\rm arg min} 
 \sum_{n \not\in \Omega} \max \big( {  \|\mathbf{W}_L\mathbf{A}_n \|_{\rm F}^2 } ~,~ {  \|\mathbf{W}_R\mathbf{A}_n^{\mathsf{T}} \|_{\rm F}^2 }\big), \nonumber\\
& 
\hspace{1cm} \quad \quad \quad  {\rm s.t.}~~
\sum_{i\in [d]} w_{L,i} = 1, ~\sum_{i \in [d']}w_{R,i} = 1.
\end{align}
%-----------
Such an optimization is a convex problem that can be tractably solved.  The solution to \eqref{eq:weightedUpper} provides us with the weight matrices used in the optimization problem \eqref{weighted-convex-noisy}.  Then, in the following subsection, we show that this weight choice reduces the weighted scenario's sample complexity to the degree of freedom of the input signal.

\subsection{Sample Complexity}

Sample complexity—the minimum number of measurements needed to ensure accurate signal reconstruction—directly~\cite{bah2016sample}. Indeed, given an array of size $N$, the sample complexity measures the sufficient number of measurements that can guarantee a perfect recovery, which affects hardware cost and acquisition time in practical DOA estimation. 

As mentioned earlier, \cite{BOKAEI2023109253} shows that for a general weighted scenario, the required samples for perfect recovery is  $\mathcal{O}(K^2\log(N))$. Here, we show that the proposed design strategy reduces the leverage scores such that the overall sample complexity becomes $\mathcal{O}(K\log(N))$.  Hence, we have the following results regarding leverage scores of the solution to \eqref{eq:weightedUpper}.

\begin{lem}\label{prop:Lev_score}
The optimal weight matrices $\mathbf{W}_L^*,~\mathbf{W}_R^*$ solution to optimization \eqref{eq:weightedUpper} make the leverage scores bounded above by a positive constant $b >4$, i.e., 
% ===============
\begin{align}
\label{diag_modified_lev_scores}
\sum_{n \in [N]} \frac{\tilde{\mu}_{n}\tilde{K}}{N} \leq  b.
\end{align}
% =============== 
\end{lem}
\begin{proof}
    See Appendix \ref{Lem:OptmialWeight}. 
\end{proof}
%----------
Lemma~\ref{prop:Lev_score} states that the proposed choice of weight matrices can reduce the leverage scores, consequently, reducing the sample complexity for the DOA recovery. In what follows, we provide the sample complexity for the weighted scenario in \eqref{weighted-convex-noisy}.
% ======================
\begin{prop}
\label{th:recovery}
 Let $\mathbf{y}\in\mathbb{C}^{N}$ be a sample vector with elements as in \eqref{eq:measured_Discrete}, and let $\Omega$ represent the set of $M$ array element locations  formed by placing an  element at location $n\in[N]$  with probability 
 %------------
 \begin{align}
  p_n =\min\{1, c{\tilde{\mu}}_{n} \Tilde{K}^2 \log^3(N)/N\},     
 \end{align}
 %------------
 independent of other locations, where $\tilde{\mu}_{n}$ is the weighted leverage score in \eqref{eq:mutildDef}. Further, let  $\tilde{\mathbf{y}}_{\Omega} = \mathcal{P}_{\Omega}(\tilde{\mathbf{y}})$  be the vector of observed noisy samples where the noise term $\mathbf{e}\in \mathbb{C}^N$  satisfies $\|\mathcal{P}_{\Omega}(\mathbf{e})\|_2\leq \sqrt{M} \eta$. Then, any solution $\widehat{\mathbf{y}}$ of \eqref{weighted-convex-noisy}, satisfies 
 %------------
 \begin{align}
     \big\|\mathbf{W}_L^*\big(\mathscr{H}(\widehat{\mathbf{y}}) -\mathscr{H}(\mathbf{y})\big) \mathbf{W}_R^{*\mathsf{T}}\big\|_{\rm F} \leq c_2\sqrt{M} \eta \frac{N}{\min_{n}p_n^2},
 \end{align}
 %------------
 for weight matrices $\mathbf{W}_L^*$ and $\mathbf{W}_R^*$ from \eqref{eq:weightedUpper}, with  probability no less than $1-N^{3-b_1}$ if
 %-------------
\begin{align}\label{eq:sample_com_exact}
M \geq c \Tilde{K}\log^4{(N)},
\end{align}
%--------------
and 
%-------------
\begin{align*}
\gamma_N\leq\min_{n\in [N]}\Big\{ \|\mathbf{A}_{n}\|_{0}\min\{\|\mathcal{P}_{U}(\mathbf{A}_{n})\|_{\rm F}^2, \|\mathcal{P}_{V}(\mathbf{A}_{n})\|_{\rm F}^2\} \Big\},
\end{align*}
%-------------
where $d$, $(N-d+1)$, and $\tilde{K}$ are the number of rows, the number of columns, and the rank of the lifted structure $\mathscr{H}(\mathbf{y})$, respectively. Additionally, $\gamma_N:=1/8\sqrt{\log(N)}$, $c=  36864(b_1+1)$ for $b_1\geq 4$ and $c_2 < 102$. 
\end{prop} 
% ======================

\begin{proof}
From  \cite[Corollary 1]{BOKAEI2023109253}, we know that for lifted structure parameter $\mathsf{R}_{\mathscr{L}}$ becomes $\log{(N)}$. Now, by invoking the obtained upper-bound in \eqref{diag_modified_lev_scores} for the leverage scores and by applying \cite[Theorem 1]{BOKAEI2023109253} to the optimization problem in \eqref{weighted-convex-noisy}, we achieve the lower-bound on the sample complexity in~\eqref{eq:sample_com_exact}. 
\end{proof}

Interestingly, Proposition~\ref{th:recovery} provides an upper bound for the sample complexity independent of the leverage scores when using the weight matrices $\mathbf{W}_{L}^{*}$ and $\mathbf{W}_{R}^{*}$. This bound is of $\mathcal{O}(\Tilde{K}\log^4{(N)})$, which is optimal up to a logarithmic factor.

% =====================
\begin{algorithm}[tb]
\caption{Lifted interpolation using ADMM}\label{alg:Low-rank-recovery}
\begin{algorithmic}[1]
	\State \textbf{Input:}
	\State {\quad Observed samples $\mathbf{y}_{\Omega}\in \mathbb{C}^{M}$ and corresponding coordinates $ \Omega \subset [N] $},
	\State {\quad Parameters ${\rho, \gamma}$ for the augmented Lagrangian form }
	\State \quad An upper-bound $\eta$ on the noise energy,
	\State \quad The number of iterations $I$,
	\State \textbf{Output:}
	\State {\quad Completed Vector }$\widehat{{\mathbf y}} \in \mathbb{C}^{N}$.
	\Procedure{Lifted interpolation}{$\mathbf{y}_{\Omega},{\rho}, {\gamma}, {\eta}$}
	\State compute $ \mathbf{W}_{L} $ and $ \mathbf{W}_{R} $ by solving \eqref{eq:weightedUpper},
	\State $\boldsymbol{\Lambda}^{(0)}\gets \mathbf{0}$,
	\State ${\mathbf{g}}^{(0)}\gets \mathcal{P}_{\Omega}{\left(\mathbf{y}\right)} $,
	\State  $\mathbf{ S}^{(0)},\mathbf{R}^{(0)} \gets \text{Polar decomposition of }\mathscr{H}(\mathbf{g}^{(0)})$
	\For {$i=1:I$}
        \State  $ \mathbf{g}^{(i)} \gets   \mathscr{H}^{\dagger} \big( \mathbf{W}_{L}^{-1}\big(\mathbf{S}^{(i)}\mathbf{R}^{(i){\mathsf{H}}}-\boldsymbol{\Lambda}^{(i)} \big)\mathbf{W}_{R}^{-{\mathsf{T} }} \big) +\mathcal{P}_{\Omega}(\mathbf{z})$
	\State $ \mathbf{F} \gets \rho \big( \mathbf{W}_{L} \mathscr{H}(\mathbf{g}^{(i)}) \mathbf{W}_{R}^{\mathsf{T} } +\boldsymbol{\Lambda}^{(i-1)}\big)$
	\State $\mathbf{S}^{(i)} \gets  \mathbf{F} \mathbf{R}^{(i-1)} \big( \mathbf{I}_{\tilde{K}}  +\rho\mathbf{R}^{(i-1)\mathsf{H}}\mathbf{R}^{(i-1)}\big)^{-1} $
	\State $\mathbf{R}^{(i)} \gets \mathbf{F}^{\mathsf{H}} \mathbf{S}^{(i)} \big( \mathbf{I}_{\tilde{K}} +\rho\mathbf{S}^{(i)\mathsf{H}}\mathbf{S}^{(i)}\big)^{-1} $
	\State $ \boldsymbol{\Lambda}^{(i)} \gets \boldsymbol{\Lambda}^{(i-1)}  +\mathscr{H}(\mathbf{g}^{(i)}) - \mathbf{S}^{(i)}{\mathbf{R}^{(i)}}^{\mathsf{H}}$
	\EndFor 
	\State $ \widehat{\mathbf{y}} =  {\mathbf g}^{({i})}$
	\State \textbf{return} $\quad\widehat{\mathbf{y}}$
	\EndProcedure
\end{algorithmic}
\end{algorithm}

% =====================

\subsection{A Fast Optimization Algorithm}

Due to the inherently large dimension of the ULA estimation problem \eqref{weighted-convex-noisy}, numerical methods for the optimization in \eqref{weighted-convex-noisy}  are often not tractable as the computational complexity is prohibitive. 
For this reason, in this subsection, we provide an SVD-free method for solving the weighted nuclear norm minimization~\eqref{weighted-convex-noisy}. In particular, the nuclear norm can be defined as \cite{srebro2004learning}:
% ===============
\begin{align}
\| \mathbf{X} \|_{\rm *} = \underset{\substack{\mathbf{S},\mathbf{R}, \mathbf{X}=\mathbf{SR}^{\mathsf{H}}}}{\min}\| \mathbf{S} \|_{\rm F}^2+\| \mathbf{R} \|_{\rm F}^2.
\label{eq:min SR}
\end{align} 
% ===============
Hence, \eqref{weighted-convex-noisy} can be reformulated as
% ============
\begin{equation}
\label{eq:minUV}
\begin{aligned}
\underset{ \substack{\mathbf{S},\mathbf{R}, \mathbf{g} \in \mathbb{C}^N}}{\rm minimize}&
& & \| \mathbf{S} \|_{\rm F}^2+\| \mathbf{R} \|_{\rm F}^2, \\
{\rm s.t.}~~~&
& & \|\mathcal{P}_{\Omega}(\mathbf{g}) - \mathcal{P}_{\Omega}( {\mathbf{y}})\|_2^2\leq \sqrt{M}\eta , \\ 
&& &\mathbf{W}_{L}\mathscr{H}(\mathbf{g})\mathbf{W}_{R}^{\mathsf{T}}  =\mathbf{SR}^{\mathsf{H}}.\\
\end{aligned}
\end{equation}
% ============
We wish to apply ADMM to this modified problem formulation. To do so, we first incorporate the constraints into the cost function as
% ============
\begin{align}
&\widetilde{L}_{\rho,\gamma}(\mathbf{S},\mathbf{R},\mathbf{g},\boldsymbol{\Lambda})=\| \mathbf{S} \|_{\rm F}^2+\| \mathbf{R} \|_{\rm F}^2+
\label{eq:admm1}
\\& \rho\|\mathbf{W}_{L}\mathscr{H}(\mathbf{g})\mathbf{W}_{R}^{\mathsf{T} } -\mathbf{SR}^{\mathsf{H}} + \boldsymbol{\Lambda} \|_{\rm F}^2   +\gamma\|\mathcal{P}_{\Omega}( \mathbf{g})  -\tilde{\mathbf{y}}_{\Omega}  \|_{\rm F}^2.\nonumber
\end{align}
% ============
In \eqref{eq:admm1}, the Lagrange multiplier $\boldsymbol{\Lambda}$ has the same size as $\mathscr{H}(\mathbf{g})$, and $\rho$ and $\gamma$ are arbitrary positive scalars.  $\gamma$  should be properly tuned to ensure $\|\mathcal{P}_{\Omega}(\mathbf{g}) -\tilde{\mathbf{y}}_{\Omega}\|_{\rm F}^2\leq {M} \eta^2$ is satisfied. We have obtained suitable $\gamma$ values through simulation results.

Note that \eqref{eq:minUV} is bi-linear in terms of $\mathbf{S}$ and $\mathbf{R}$, but not necessarily convex. As shown in   \cite{ye2016compressive,hong2015convergence},  the above ADMM method is known to converge when the penalty parameter $\rho$ is sufficiently large.  For the ADMM formulation, we divide the minimization in \eqref{eq:min SR} into simpler sub-problems. 
Let $\mathbf{g}^{(i)}$,$\mathbf{S}^{(i)}$, $\mathbf{R}^{(i)}$, and $\boldsymbol{\Lambda}^{(i)}$  be the matrices estimates at the $i$-th iteration. The update rules for the matrix $\mathbf{g}^{(i+1)}$ is
% ============
\begin{align}
\label{eq.gUpdate1}
\mathbf{g}^{(i+1)} =  & \mathscr{H}^{\dagger} \big( \mathbf{W}_{L}^{-1}\big(\mathbf{S}^{(i)}\mathbf{R}^{(i){\mathsf{H}}}-\boldsymbol{\Lambda}^{(i)} \big)\mathbf{W}_{R}^{-{\mathsf{T} }} \big) +\mathcal{P}_{\Omega}(\mathbf{z}),
\end{align}
% ============
where the  elements of $\mathbf{z} \in \mathbb{C}^{N}$  are given by
% ============
\begin{align}
\mathbf{z}_j & = \Big(\gamma\mathbf{y}_j + \rho~\Big[\mathscr{H}^{\dagger} \big(\mathbf{S}^{(i)}\mathbf{R}^{(i)} - \boldsymbol{\Lambda}^{(i)} \big)\Big]_j\Big) \nonumber
\\ 
& \quad \cdot \Big(\gamma + \rho 
\Big[\mathscr{H}^{\dagger} \mathscr{H}(\mathbf{e}_{j})\Big]_j\Big)^{-1}, 
\end{align}
% ============
in which $[\cdot]_j$ refers to the $j$-th element of the vector. Then, for matrices $\mathbf{S}^{(i+1)}$ and $\mathbf{R}^{(i+1)}$, we have	
% ============
\begin{align}
\nonumber
& \mathbf{S}^{(i+1)}  \\
& = \mathrm{arg}\min_{\mathbf{S}}\| \mathbf{S} \|_{\rm F}^2 + \rho\|\mathbf{W}_{L}\mathscr{H}(\mathbf{g}^{(i+1)})\mathbf{W}_{R}^{\mathsf{T} } -\mathbf{SR}^{(i){\mathsf{H}}}+\boldsymbol{\Lambda}^{(i)} \|_{\rm F}^2,   \nonumber\\ \label{eq:updateS2}
& = \rho\big(\mathbf{W}_{L}\mathscr{H}(\mathbf{g}^{(i+1)})\mathbf{W}_{R}^{\mathsf{T} }
+\boldsymbol{\Lambda}^{(i)}\big)\mathbf{R}^{(i)}\cdot \big( \mathbf{I}_{\tilde{K}}  +\rho\mathbf{R}^{(i){\mathsf{H}}}\mathbf{R}^{(i)}\big)^{-1},  
\end{align}
% ============
and similarly
% ============
\begin{align}
\nonumber
& \mathbf{R}^{(i+1)}  \\
& =  \mathop{\mathrm{argmin}}_{\mathbf{R}}\| \mathbf{R} \|_{\rm F}^2 \hspace{-0.5mm}+\hspace{-0.5mm} \|\mathbf{W}_{L}\mathscr{H}(\mathbf{g}^{(i+1)})\mathbf{W}_{R}^{\mathsf{T} } \hspace{-0.5mm}-\hspace{-0.5mm} \mathbf{S}^{(i+1)}\mathbf{R}^{\mathsf{H}} \hspace{-0.5mm}+\hspace{-0.5mm} \boldsymbol{\Lambda}^{(i)} \|_{\rm F}^2, \nonumber \\
& = \rho\big(\mathbf{W}_{L}\mathscr{H}(\mathbf{g}^{(i+1)})\mathbf{W}_{R}^{\mathsf{T} }
+\boldsymbol{\Lambda}^{(i)}\big)^{\mathsf{H}}\mathbf{S}^{(i+1)} \nonumber \\
& \quad\quad \quad \cdot \big( \mathbf{I}_{\tilde{K}} +\rho\mathbf{S}^{(i+1){\mathsf{H}}}\mathbf{S}^{(i+1)}\big)^{-1}, \label{eq:updatev}
\end{align}
% ============
and the standard update rule for the Lagrangian multiplier
% ============
\begin{align}
\label{eq:lambda1}
\boldsymbol{\Lambda}^{(i+1)} = & \boldsymbol{\Lambda}^{(i)}  + \mathscr{H}(\mathbf{g}^{(i+1)}) - \mathbf{S}^{(i+1)}\mathbf{R}^{(i+1) {\mathsf{H}}}.
\end{align}
% ============
Algorithm \ref{alg:Low-rank-recovery} provides a unified summary of the overall procedure of the proposed method.   In the next subsection, we reformulate the algorithm for DOA estimation.

\subsection{Computational Complexity}

Computational complexity quantifies the number of arithmetic operations required for performing Algorithm\ref{alg:Low-rank-recovery}, directly affecting run‐time and real‐time applicability. The operation includes all elementary arithmetic operations—additions, subtractions, multiplications, and divisions. The computational complexity of the proposed algorithm primarily arises from the matrix operations in  \eqref{eq.gUpdate1}, \eqref{eq:updateS2} , and \eqref{eq:updatev}. Specifically, updating \( g^{(i+1)} \) in  \eqref{eq.gUpdate1} involves the inversion of a matrix of size \( d \times d \), leading to a complexity of \( \mathcal{O}(d^3) \). Similarly, the updates of \( S^{(i+1)} \) and \( R^{(i+1)} \) in \eqref{eq:updateS2} and \eqref{eq:updatev} require inverting matrices of size \( \tilde{K} \times \tilde{K}\) and \( d \times d \), resulting in \( \mathcal{O}(\tilde{K}^3) \) and \( \mathcal{O}(d^3) \), respectively. In addition to these matrix inversions, matrix multiplications introduce computational costs of \( \mathcal{O}(d^2 \tilde{K}) \). Overall, the per-iteration complexity of the proposed algorithm is \( \mathcal{O}(d^3 + \tilde{K}^3 + d^2 \tilde{K}) \), where \( d \) is the size of the structured matrix and \( \tilde{K} \) is the rank.

In contrast, solving the nuclear norm minimization problem in \eqref{weighted-convex-noisy} directly using CVX involves solving the SDP problem. The complexity of SDP solvers is highly dependent on the size of the lifted matrix and the number of constraints. For a structured matrix of size \( d \times (N-d+1) \), the computational complexity for each iteration of CVX is approximately \( \mathcal{O}(d^3 (N-d+1)^3) \), which is significantly higher than the proposed algorithm. Additionally, SDP solvers require storing all constraints explicitly, leading to storage complexity of \( \mathcal{O}(d^2 (N-d+1)^2) \), compared to the \( \mathcal{O}(d(N-d+1) + d^2) \) storage required by the proposed method.
Thus, the proposed ADMM-based approach significantly reduces the computational burden and storage requirements compared to CVX while achieving comparable recovery accuracy. This efficiency makes the method more suitable for real-time applications.

\section{DOA Estimation for  ULA}\label{sec:DOAestiion_ula}

We have previously described how to interpolate the measured data on a non-uniform array ($\mathbf{y}_{\Omega}$) to obtain a full uniform array ($\widehat{\mathbf{y}}$). Next, we need to estimate the DOAs. The Prony method is one of the simplest yet successful techniques for noiseless samples~\cite{prony1795essai}. Due to the involved algebraic operations, the performance substantially degrades with noise~\cite{chen2014robust}. Therefore, we opt to use a super-resolution technique to estimate the final DOAs; the involved optimizations add to the complexity of the method but make the procedure more robust to non-idealities such as additive noise.

% ================================================
\begin{figure*}[!t]
\centering

\subfigure[]{\label{fig:locations(a)}\includegraphics[width=0.24\linewidth]{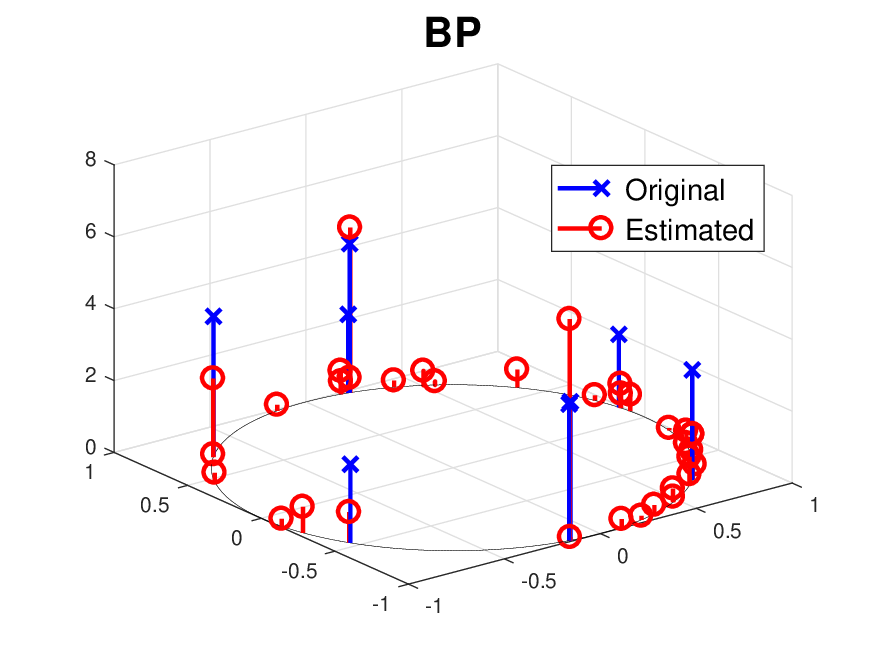}}
\subfigure[]{\includegraphics[width=0.24\linewidth]{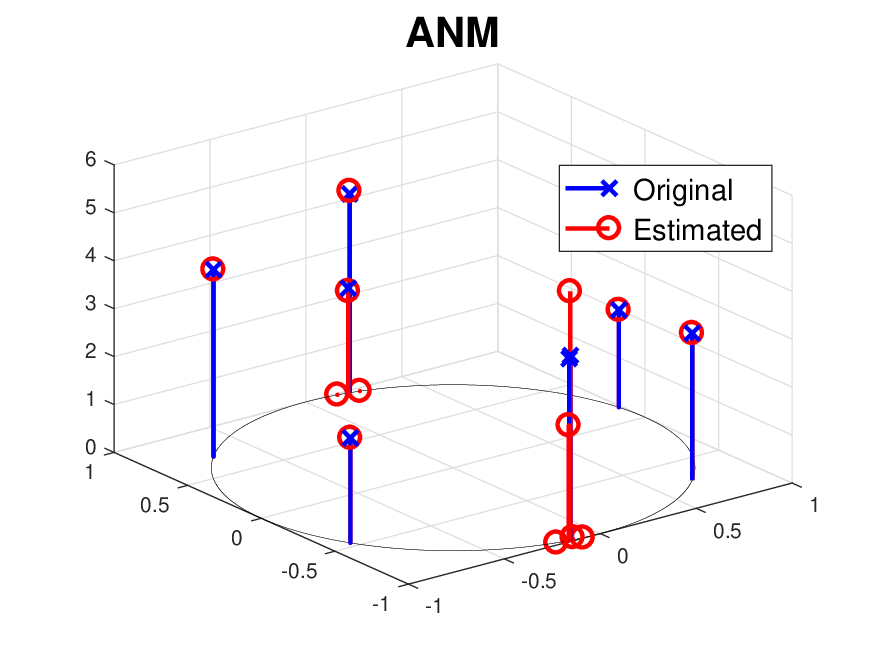}}
\subfigure[]{\includegraphics[width=0.24\linewidth]{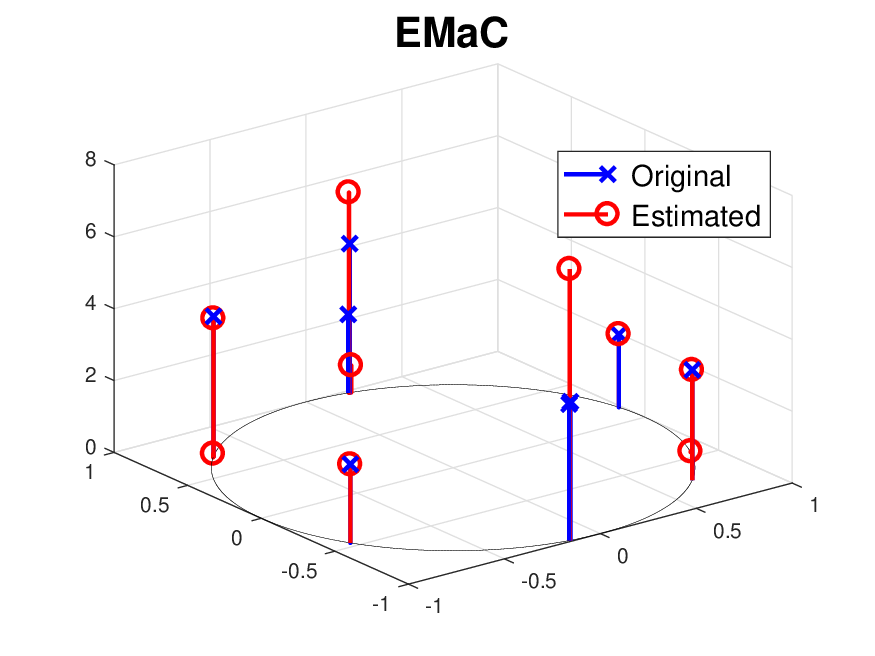}}
\subfigure[]{\label{fig:locations(d)}\includegraphics[width=0.24\linewidth]{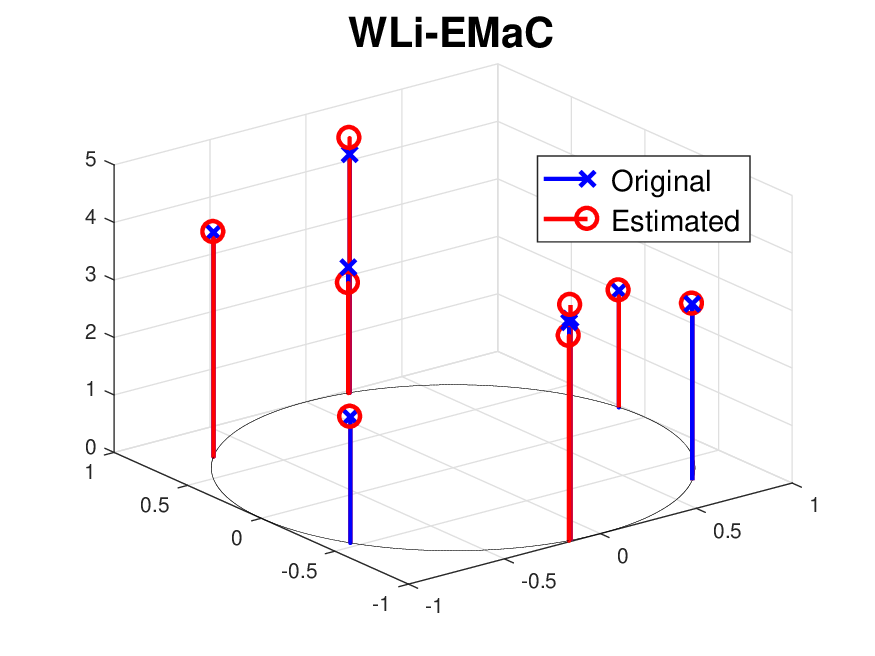}}

\subfigure[]{\label{fig:locations(e)}\includegraphics[width=0.24\linewidth]{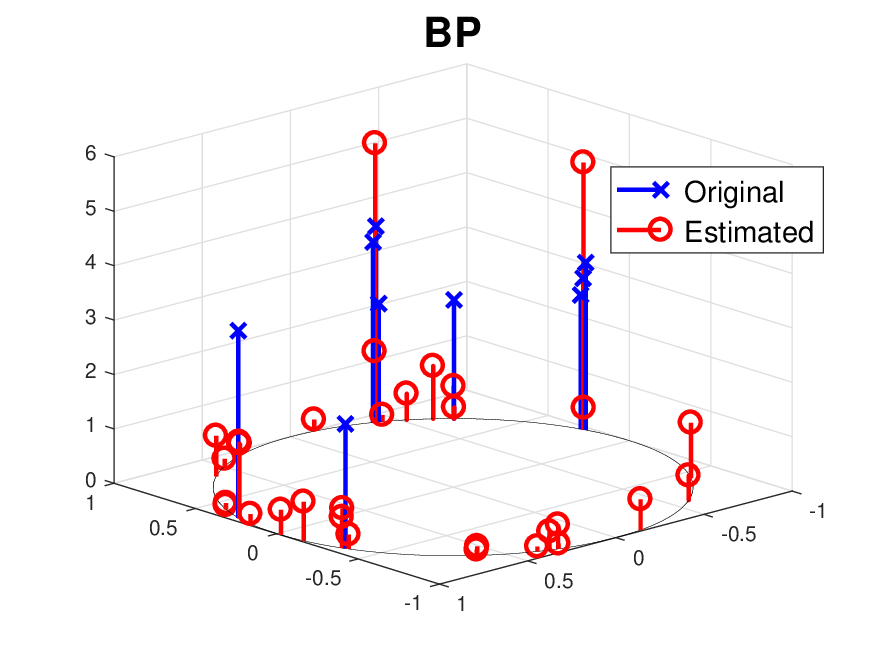}}
\subfigure[]{\includegraphics[width=0.24\linewidth]{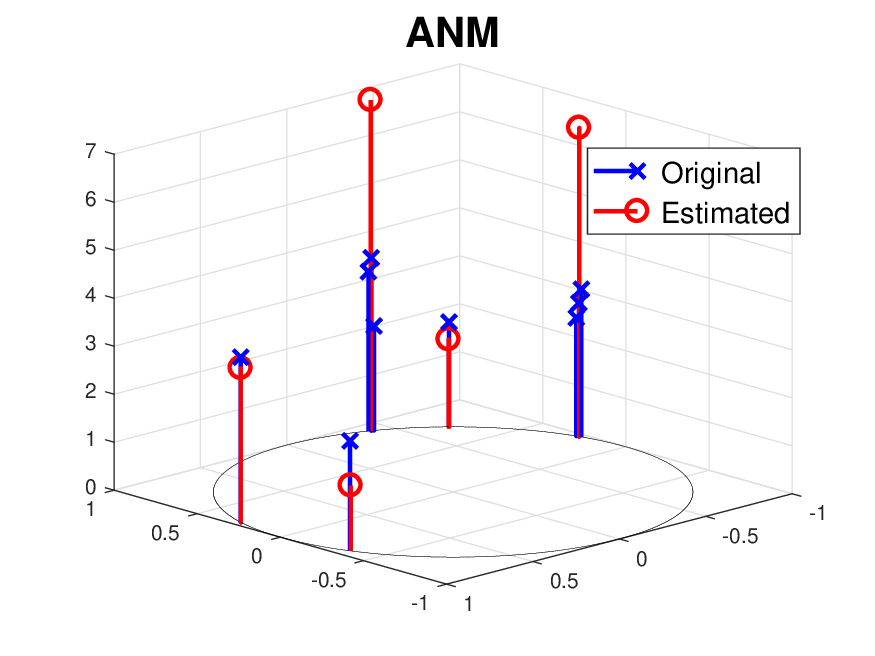}}
\subfigure[]{\includegraphics[width=0.24\linewidth]{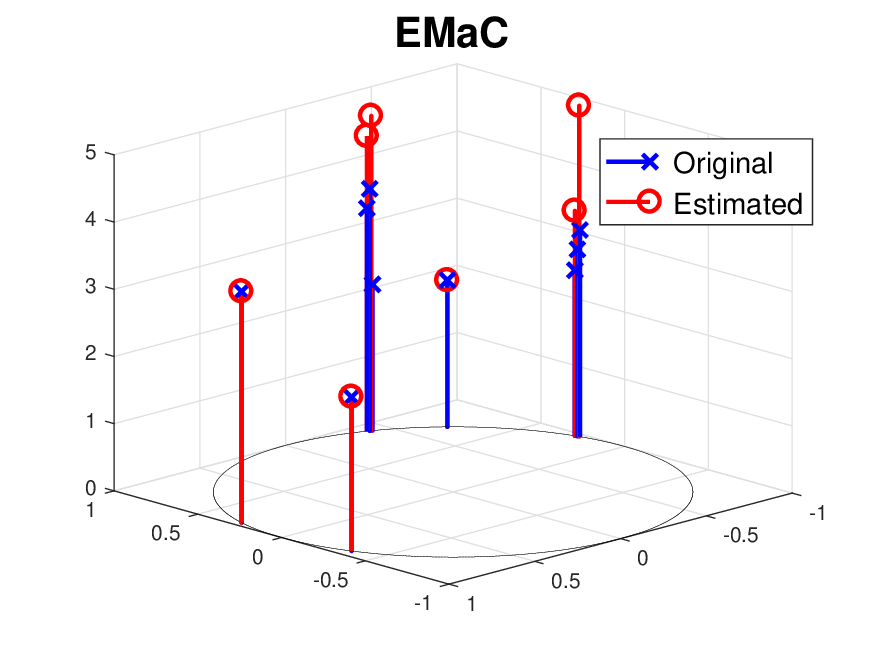}}
\subfigure[]{\label{fig:locations(h)}\includegraphics[width=0.24\linewidth]{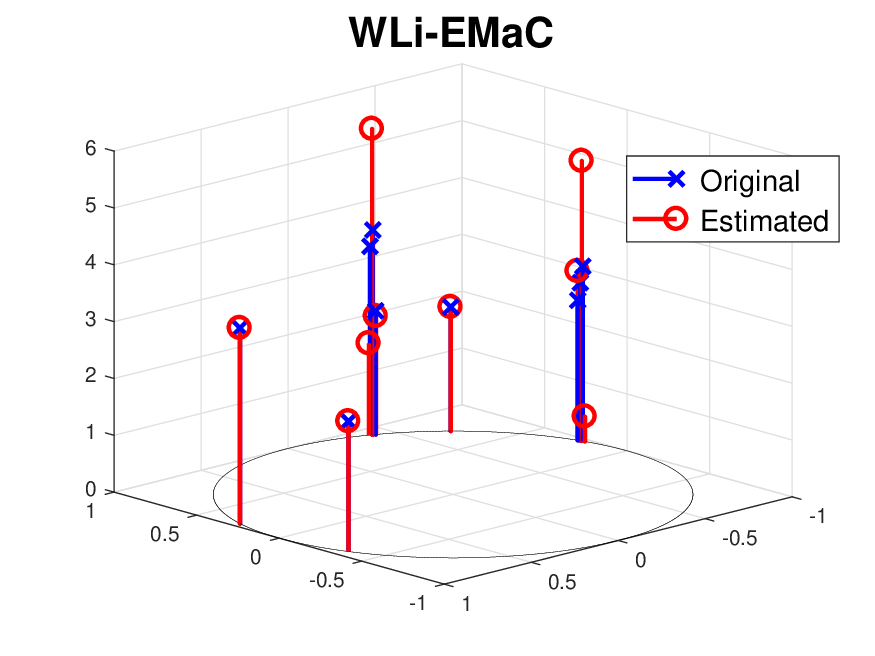}}

\caption{DOA estimation performance of BP, ANM, EMaC, and WLi-EMaC algorithms in Sec.~\ref{sec:DOA Estimation}; the circles represent the $360^{\circ}$ angular space and the height of the bars indicate the strength of the received signal from each source.  The $3$ source setups consist of $8$ different sources in the first row, two pairs of which are seemingly co-angled, and $9$ sources with two almost co-angled triplets in the second row.}
\label{fig:locations}
\end{figure*}
% ================================================

By completing the non-uniform array elements $\mathbf{y}_{\Omega}$, we use the super-resolution method to estimate DOAs from the estimated full linear uniform array, i.e., $\widehat{\mathbf{y}}$. This problem was previously studied in \cite{candes2014towards}. For completeness, we provide a summary of this method below. To recover the DOAs,  we estimate $\tau_{k}$s along with their amplitudes $b_{k}$. For this purpose, similar to \eqref{convex-TV}, we minimize the TV norm of a continuous-domain signal $h$ that produces  $\widehat{\mathbf{y}}$ in a linear fashion:
% ============
\begin{align}
\label{convex-TV-2}
\underset{h}{\rm minimize} ~\| h \|_{\mathsf{TV}} ~~~{\rm s.t.}~ \mathcal{A}(h) = \widehat{\mathbf{y}},
\end{align}
% ============
where $\|h\|_{\mathsf{TV}}$ is equal to the amplitude sum $\sum_{k\in [K]} b_{k}$. Here, $ \widehat{\mathbf{y}}$ stands for the recovered full linear uniform array from  Algorithm \ref{alg:Low-rank-recovery}.  Using the dual formulation, the infinite-dimensional optimization in \eqref{convex-TV-2} reduces to finding the roots of a trigonometric polynomial. More precisely, the Lagrange dual problem associated with \eqref{convex-TV-2} reads \cite{candes2014towards} as
% ---------------
\begin{align}
\label{dual}
\mathbf{q}_{\text{opt}}  = \underset{\mathbf{q} \in \mathbb{C}^{N}}{\rm argmax} ~{\mathfrak{R}}\big(\mathbf{q}^{\mathsf H}\widehat{\mathbf{y}} \big) ~~~{\rm s.t.}~~~ \underset{\tau}{\rm sup}|\mathbf{q}^{\mathsf{H}} \mathbf{a}(\tau)| \leq 1,
\end{align}
% --------------
where $\mathbf{a}(\tau)$  is  $\mathbf{a}(\tau) =\big[a_{0}(\tau),~ a_{1}(\tau), \ldots,a_{N-1}(\tau) \big]^{\mathsf{T} },$ with $a_{n}(\tau)$ following the definition in \eqref{eq:atom}. For finding $\tau_{k}$s, we shall solve $|\mathbf{q}_{\text{opt}}^{\mathsf{H}}\mathbf{a}(\tau)|=1$ (a trigonometric polynomial equation). After deriving $\tau_{k}$s, we can obtain $b_{k}$s by solving the system of equations $\sum_{k\in [K]}{\rm e}^{-{\rm j} 2 \pi \tau_{k} n}b_{k} = \widehat{\mathbf{y}}[n]$ for $n \in [N]$, using the method of least squares.

\begin{remark}
The formulation in \eqref{convex-TV-2} enforces exact data fitting and hence admits perfect recovery in the noiseless regime whenever the sample‐complexity conditions of Proposition~\ref{th:recovery} are met. In practice, however, measurements are often contaminated by noise. To account for this, one may replace the hard equality constraint in \eqref{convex-TV-2} with a tolerance ball below. 
%-----------
\begin{align}
    \|\mathcal{A}(h) - \widehat{\mathbf{y}}\|_{F} \;\le\;\kappa,
\end{align}
%-----------
where \(\kappa\) can be chosen according to the upper bound on the interpolation error. The lifted error given in Proposition~\ref{th:recovery} can also provide an upper bound for the interpolation error, i.e., $\kappa =c_2\sqrt{M} \eta \frac{N}{\min_{n}p_n^2}$.  Although this provides a principled estimate of the allowable deviation (and thus a guideline for selecting \(\kappa\)), the resulting bound is generally conservative and need not be tight in all scenarios.
\end{remark}

% ================================================
\begin{table}[t!]
\centering
\caption{  The three source setups in DOA estimation simulations for Figures \ref{fig:locations}, \ref{fig:accuracy}, \ref{fig:snracc} and \ref{fig:gamma_effect} in  Secs. \ref{sec:DOA Estimation}, \ref{sec:The Effect of The Array Size}, \ref{sec:Robustness Against Noise}, and \ref{ADMM Algorithm versus Convex Implementation} }
\begin{tabular}{ |m{0.8cm}|m{4cm}|m{2cm}| }
\hline
Scenario & Angle of Sources & Amplitudes \\ 
\hline
(a) & $-71.56^{\circ}$, $-34.58^{\circ}$, $-34.44^{\circ}$, $-14.58^{\circ}$, $ 8.74^{\circ}$, $26.80^{\circ}$, $26.92^{\circ}$, $48.08^{\circ}$ & $3.79$, $3.06$, $3.89$, $4.14$, $2.18$, $2.02$, $3.85$,$2.18$ \\
\hline 
(b) &  $-36.17^{\circ}$, $-22.66^{\circ}$, $35.20^{\circ}$, $ 35.48^{\circ}$, $35.76^{\circ}$, $49.60^{\circ}$, $59.66^{\circ}$, $60.11^{\circ}$, $60.58^{\circ}$
& $3.07$, $2.28$, $2.17$, $2.18$, $3.43$, $2.78$, $2.46$, $3.60$, $3.31$ \\
\hline 
(c) & $-25.48^{\circ}$, $-24.04^{\circ}$, $-22.79^{\circ}$, $47.12^{\circ}$, $48.83^{\circ}$ &  $2.66$, $2.59$, $2.61$, $3.75$, $2.83$\\
\hline
(d) & $-30.46^{\circ}$, $-6.89^{\circ}$, $41.29^{\circ}$ &  $2.62$, $3.86$, $3.48$\\
\hline
\end{tabular}
\label{tab:source_setups}
\end{table}
% ================================================

\subsection{DOA Recovery Performance}

The central issue revolves around the degree to which ULA estimation can improve the DOA estimation problem in \eqref{dual}. To address this matter, we need to identify the conditions under which the estimated $\hat{\theta}_k$s coincide with the true $\hat{\theta}_k$s. In particular, the recovery of DOAs depends on the minimal separation of DOAs $\{\theta_k\}_{k=1}^{K}$ or the minimal wrap-around distance between any pair of distinct DOAs.

% ---------------
\begin{prop}(\cite[Theorem 2.2]{fernandez2016super})\label{prop:DOA_bound}
    Let $\Theta = \{ \theta_k \}_{k=1}^K$ be the set be all the received DOAs in \eqref{eq:measurement all} for an array of size $N$ and the $\Delta (\Theta)$ be the minimal wrap-around distance defined as
    \begin{align}
         \Delta(\Theta) = \inf_{\substack{ \theta_i, \theta_j \in \Theta \\\theta_i \neq \theta_j}} \min_{q \in \mathbb{Z}}\big|\sin(\theta_i) - \sin(\theta_j) + q\big|. 
     \end{align}
    For $N\gg 1$, if the minimum separation obeys 
     \begin{align}
         \Delta(\Theta)\geq \frac{\lambda 2.52}{s_d N}, 
     \end{align}
     then $\mathbf{q}_{\text{opt}}$ is the unique solution to \eqref{convex-TV-2}. Here, $s_d$ denotes the spacing of array elements, and $\lambda$ is the wavelength of the reflected signal from the source. 
\end{prop}
% ---------------
Proposition~\ref{prop:DOA_bound} clearly indicates that for a ULA with $N'$ antenna elements, the separation condition must be $\Delta(\Theta) \leq \mathcal{O}(1/N')$. Therefore, by using the proposed weighted method alongside Proposition~\ref{th:recovery}, it is deducible that, with a high probability, the separation condition $\Delta(\Theta)$ for SLA array of size $M = cK\log^4(N)$,  enhances from an initial value $\mathcal{O}(1/K\log^4(N))$ to $\mathcal{O}(1/N)$.

\begin{remark}
We note that both the denoising and interpolation steps rely on the gridless recovery principle, which implies that DOAs can be estimated without prior knowledge of the number of sources.  Moreover, although measurements are corrupted by Gaussian noise, Proposition~\ref{th:recovery} shows that the interpolation error decays on the order of \(1/N\), so that for sufficiently large \(N\) and moderate to high SNR the recovered lifted matrix converges arbitrarily close to the true one, which in turn guarantees accurate DOA estimation without prior knowledge of the number of sources.
\end{remark}

\section{Generalization to Multiple Snapshot Case}

Extending the proposed DOA estimation framework to a multiple-snapshot scenario significantly enhances the robustness and applicability of our approach. To this end, the generalization is addressed through three principal steps: First, we generalize the Hankel transform to multiple snapshots; second, we adapt the design of the weight matrices to account for the aggregated structure across snapshots; and third, we provide a method for performing estimation using these extended weights.

More precisely, consider a set of $T$ measurement vectors (snapshots), each denoted by $\mathbf{y}_t \in \mathbb{C}^{N}$, with $t \in [T]$. Each vector $\mathbf{x}_t$ retains a common spectral structure, permitting a block Hankel matrix representation. Thus, an extended block Hankel matrix $\mathscr{H}_{d_t,d}: \mathbb{C}^{N\times T}\mapsto \mathbb{C}^{d_td\times(T-d_t+1)(N-d+1)}$ is defined as:
%---------------
\begin{equation*}
\label{eq:Hankel}
\mathcal{H}_{d_t,d}(\mathbf{Y}):= \left[ \begin{matrix}
\mathscr{H}(\mathbf{y}_1) & \mathscr{H}(\bm{y}_2) & \dots& \mathscr{H}(\bm{y}_{T-d_t+1})\\
\mathscr{H}(\bm{y}_2) & \mathscr{H}(\bm{y}_3) & \dots& \mathscr{H}(\bm{y}_{T-d_t+2})\\
\vdots & \vdots & \ddots&  \vdots\\
\mathscr{H}(\bm{y}_{d_t}) & \mathscr{H}(\bm{y}_{d_t+1}) & \dots& \mathscr{H}(\mathbf{y}_{T})
\end{matrix}  \right],
\end{equation*}
%---------------
where $\mathbf{Y}:=[\mathbf{y}_1,\ldots,\mathbf{y}_T]\in \mathbb{C}^{N\times T}$ contains all $T$ snapshot measurements and each $\mathscr{H}(\mathbf{y}_t)\in \mathbb{C}^{d\times (N-d+1)}$ is the Hankel transform of snapshot $t$. This concatenation exploits spectral correlations across snapshots, leading to a low-rank structure whose rank is bounded by the number of distinct spectral components shared across the snapshots~\cite{Chen2013SpecteralHankel,razavikia2019reconstruction}. Consequently, we interpolate the matrix of $T$ multiple measurements by solving the following weighted matrix completion. 
% ===============
\begin{align}
\nonumber
\widehat{\mathbf{Y}} = & \underset{ \mathbf{G} \in \mathbb{C}^{N\times T}}{\rm argmin}~\| \mathbf{W}_L^{b}\mathcal{H}({\mathbf{G}})\mathbf{W_R}^{b,{\mathsf{H}} }\|_{*},~~ \\ &{\rm s.t.}~~\| \mathcal{P}_{\Omega}( \mathbf{G} ) -\mathcal{P}_{\Omega}(\mathbf{Y}) \|_{\rm F} \leq {\sqrt{TM}}\eta, \label{block-weighted-convex-noisy}
\end{align}
% ===============
where $\mathbf{W}_L^{b}$ and $\mathbf{W}_R^{b}$ are left and right square weighed matrices with size of $d_td\times d_td$ and $(T-d_t+1)(N-d+1)\times (T-d_t+1)(N-d+1)$, respectively. While for the interpolation step, we do not need to impose any structure on the weighted matrices, and the leverage score can be refined by only replacing the weighted matrices $\mathbf{W}_L^{b}$ and $\mathbf{W}_R^{b}$ in \eqref{eq:mutildDef}, we incorporate the concatenated Hankel structure for design the weighted matrices to use Proposition~\ref{prop:Lev_score}. Indeed,  we consider the following block Hankel structure for the weighted matrices.
%---------------
\begin{equation*}
\label{eq:Hankel_new}
\mathscr{H}_{d}(\mathbf{W}):= \left[ \begin{matrix}
\mathbf{W}_L^{(1)}& \mathbf{W}_L^{(2)} & \dots& \mathbf{W}_L^{(T-d_t+1)}\\
\mathbf{W}_L^{(2)} & \mathbf{W}_L^{(3)} & \dots& \mathbf{W}_L^{(T-d_t+2)}\\
\vdots & \vdots & \ddots&  \vdots\\
\mathbf{W}_L^{(d_t)} & \mathbf{W}_L^{(d_t+1)} & \dots& \mathbf{W}_L^{(T)}
\end{matrix}  \right], 
\end{equation*}
%---------------
where each block of  $\mathbf{W}_{L}^{(t)}$ or $\mathbf{W}_{R}^{(t)}$ are restricted to be diagonal as in \eqref{eq:wighetddiag}. Then, to design the weighted matrix, we pose the following minimization,

%------------
\begin{align}
\label{weight_cald}
\{w_{L,i}^{(t)}\}_{i\in [d]},~\{w_{R,i}^{(t)}\}_{i\in [d'] } =  \underset{w_{L,i}^{(t)}, w_{R,i}^{(t)} \in \mathbb{R}_+ }{\rm argmin} 
\sum_{n \not\in \Omega} \tilde{\mu}_{n}, 
\end{align}
%------------
where $w_{L,i}^{(t)}$ denotes element $i$ of diag of the weight matrix $\mathbf{W}_{L}^{(t)}$. Similarly, to obtain a unique minimizer for the cost in \eqref{weight_cald}, we propose  the sum of
each diag weight to be a constant, i.e, 
%------------
\begin{align}
\nonumber
& \{w_{L,i}^{(t)*}\}_{i\in [d]},~\{w_{R,i}^{(t)*}\}_{i \in [d']}    
= \underset{w_{L,i}^{(t)}, w_{R,i}^{(t)} \in \mathbb{R}_+ }{\rm argmin} 
\sum_{n \not\in \Omega} \mu_{n}, \nonumber \\ 
& \qquad \qquad \qquad  \text{s.t.}~ \sum_{i\in [d]} w_{L,i}^{(t)} = 1, ~\sum_{i \in [d'] }w_{R,i}^{(t)} = 1.\label{eq:weight_cal_3_new}
\end{align}
%------------
The advantage of such a block Hankel structure for the weight matrices is that we can use similar arguments from \eqref{eq:weight_cal_3} to \eqref{eq:weightedUpper} and the fact that the constraints are over each block, and 
optimizing each weight block,  $\mathbf{W}_L^{(t)}$ and $\mathbf{W}_R^{(t)}$ separately. Indeed, the upper bound on $\mu_n$ is the Frobenius norm and decomposable in terms of each block, and together with separate constraints, makes the optimization of the weighted separable. Specifically, each block of diagonal weights can be obtained by solving the following $T$ optimizations. 
\begin{align}
\nonumber
\mathbf{W}_L^{(t)*}, \mathbf{W}_R^{(t)*} \hspace{-4pt}&= \hspace{-3pt}\underset{\mathbf{W}_L^{(t)}, \mathbf{W}_R^{(t)}}{\mathrm{argmin}}\hspace{-2pt}\sum_{n\not\in\Omega^{(t)}}\max\left(|\mathbf{W}_L^{(t)}\mathbf{A}_n|_{\mathrm{F}}^2, |\mathbf{W}_R^{(t)}\mathbf{A}_n^{\mathsf{T}}|_{\mathrm{F}}^2\right),\\
&\text{s.t.}\quad \sum_{i\in[d]} w_{L,i}^{(t)}=1,\quad \sum_{i\in[d']} w_{R,i}^{(t)}=1, \label{eq:eighteblock}
\end{align}
where $\Omega^{(t)}$ denotes the subset of the antenna selected at snapshot $t$. We note that in most cases $\Omega^{(t)}$ is the same for all the snapshots, as we do not change the location of the antenna array in most of the applications. Consequently, all $T$ optimizations would result in the same weighted matrices. After solving these $T$ convex optimizations in \eqref{eq:eighteblock}, we complete the array over multiple snapshots by solving the weighted optimization problem in \eqref{block-weighted-convex-noisy}.

% ================================================
\begin{figure*}[!t]
\centering
\subfigure[]{\label{fig:accuracy(a)}
\begin{tikzpicture} 
    \begin{axis}[
        xlabel={ Number of Samples},
        ylabel={Recovery},
        label style={font=\footnotesize},
         ticklabel style = {font=\footnotesize},
        %legend cell align={left},
        width=0.49\textwidth,
        height=5.5cm,
        xmin=5, xmax=26,
        legend style={nodes={scale=0.6, transform shape}, at={(0.3,0.99)}}, 
                minor tick num=5,
        ymajorgrids=true,
        xmajorgrids=true,
        grid=both,
        grid style={line width=.1pt, draw=gray!15},
        major grid style={line width=.2pt, draw=gray!40},
    ]
    \addplot[
        color=cadmiumgreen,
        mark=square,
        line width=1pt,
        mark size=2pt,
        ]
    table[x=nSamp,y=WDEMaC]
    {data/Error_5.dat};
    \addplot[
        color=cadmiumorange,
        mark=square,
        mark options = {rotate = 45},
        line width=1pt,
        mark size=2pt,
        ]
    table[x=nSamp,y=DEMaC]
    {data/Error_5.dat};
    \addplot[
        color=darkcerulean,
        mark=triangle,
        line width=1pt,
        mark size=2pt,
        ]
    table[x=nSamp,y=WEMaC]
    {data/Error_5.dat};
    \addplot[
        color=darklavender,
        mark=triangle,
        mark options = {rotate = 180},
        line width=1pt,
        mark size=2pt,
        ]
    table[x=nSamp,y=EMaC]
    {data/Error_5.dat};
    \addplot[
        color=jazzberryjam,
        mark=star,
        line width=1pt,
        mark size=3pt,
        ]
    table[x=nSamp,y=ANM]
    {data/Error_5.dat};
    \addplot[
        color=bazaar,
        mark=o,
        line width=1pt,
        mark size=2pt,
        ]
    table[x=nSamp,y=BP]
    {data/Error_5.dat};
    \legend{WLi-DEMaC, DEMaC, WLi-EMaC, EMaC, ANM, BP};
    \end{axis}
    \end{tikzpicture}
}
\subfigure[]{\label{fig:accuracy(b)}
\begin{tikzpicture} 
    \begin{semilogyaxis}[
        xlabel={Number of Samples},
        ylabel={ NMSE},
        label style={font=\footnotesize},
        %legend cell align={left},
        width=0.49\textwidth,
        height=5.5cm,
        xmin=5, xmax=26,
        legend style={nodes={scale=0.65, transform shape}, at={(0.45,0.65)}}, 
        ticklabel style = {font=\footnotesize},
        minor tick num=5,
        ymajorgrids=true,
        xmajorgrids=true,
        grid=both,
        grid style={line width=.1pt, draw=gray!15},
        major grid style={line width=.2pt, draw=gray!40},
        %legend pos=east,
        %ymajorgrids=true,
        %xmajorgrids=true,
        %grid style=dashed,
         ymode = log,
        grid=both,
        grid style={line width=.1pt, draw=gray!10},
        major grid style={line width=.2pt,draw=gray!30},
    ]
    \addplot[
        color=jazzberryjam,
        mark=star,
        line width=1pt,
        mark size=3pt,
        ]
    table[x=nSamp,y=ANM]
    {data/NMSE_5.dat};
    \addplot[
        color=darklavender,
        mark=triangle,
        mark options = {rotate = 180},
        line width=1pt,
        mark size=2pt,
        ]
    table[x=nSamp,y=EMaC]
    {data/NMSE_5.dat};
    \addplot[
        color=darkcerulean,
        mark=triangle,
        line width=1pt,
        mark size=2pt,
        ]
    table[x=nSamp,y=WEMaC]
    {data/NMSE_5.dat};
    \addplot[
        color=cadmiumorange,
        mark=square,
        mark options = {rotate = 45},
        line width=1pt,
        mark size=2pt,
        ]
    table[x=nSamp,y=DEMaC]
    {data/NMSE_5.dat};
    \addplot[
        color=cadmiumgreen,
        mark=square,
        line width=1pt,
        mark size=2pt,
        ]
    table[x=nSamp,y=WDEMaC]
    {data/NMSE_5.dat};
    \legend{ANM, EMaC, WLi-EMaC, DEMaC, WLi-DEMaC};
    \end{semilogyaxis}
    \end{tikzpicture}
}
\caption{ Scenario in  Sec. \ref{sec:The Effect of The Array Size}.(a) and (b) respectively represent the probability of correct DOA estimation (BP, Atomic, EMaC, DEMaC, WLi-EMaC and WLi-DEMaC algorithms) and the normalized mean square error of the SLA interpolation to form the ULA (Atomic, EMaC, DEMaC, WLi-EMaC algorithms) in terms of the size of the SLA for a fixed source setup with $5$ sources, two and triplet of which are almost co-angled.
}
\label{fig:accuracy}
\end{figure*}
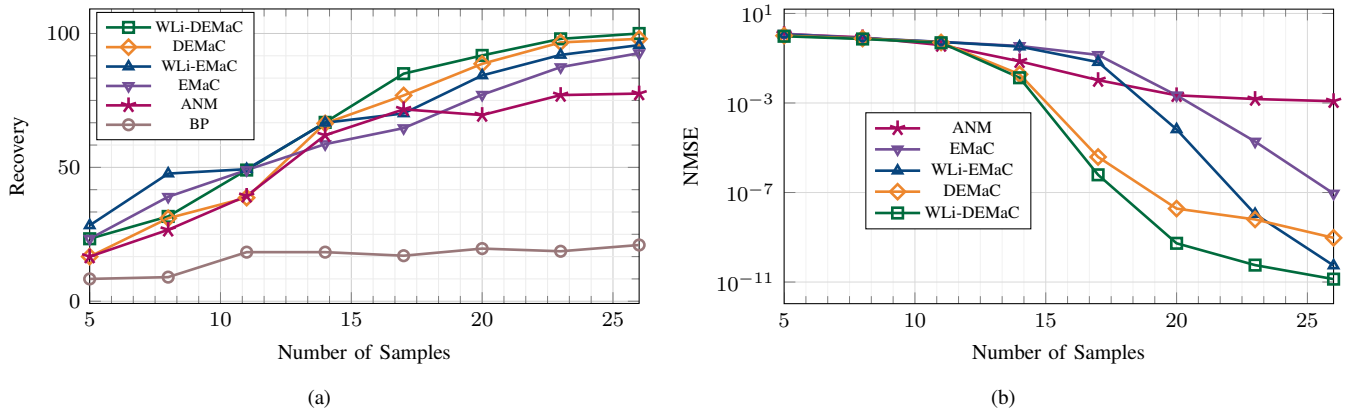
% ================================================

By completing the non‐uniform array elements $\mathbf{Y}_\Omega$, we stack the $T$ estimated ULA vectors into $\widehat{\mathbf{Y}}\in\mathbb{C}^{N\times T}$ and recover the standard spectral components by solving the multiple measurement vector atomic‐norm method~\cite{li2015off}. In the noiseless case, we enforce exact data fitting:
\[
\widehat{\mathbf{X}}
=\arg\min_{\mathbf{X}\in\mathbb{C}^{N\times T}}
\|\mathbf{X}\|_{\mathcal{A}}
\quad\text{s.t.}\quad
\mathcal{A}(\mathbf{X})=\widehat{\mathbf{Y}},
\]
where the MMV atomic norm is defined via the continuous dictionary of rank‐one atoms 
\begin{align}
    \mathcal{A} = \bigl\{\,\mathbf{a}(\tau)\,\mathbf{b}^H:\,\tau\in[0,1),\;\|\mathbf{b}\|_2=1,\mathbf{b}\in \mathbb{C}^{T}\bigr\},
\end{align}
with $\bm{a}(\tau) :=[a_0(\tau),\ldots,a_N(\tau)]$  and 
\begin{align*}
    \|\mathbf{X}\|_{\mathcal{A}}
& =\inf\Bigl\{\sum_kc_k:\mathbf{X}=\sum_kc_k\,\mathbf{a}(\tau_k)\mathbf{b}_k^{\mathsf{H}},\\&~~~~~~~~~~~c_k\ge0,\;\|\mathbf{b}_k\|_2=1\Bigr\}.
\end{align*}
 Both programs admit equivalent semidefinite‐program representations and can be solved efficiently. The dual problem of the noiseless formulation reads
\begin{align*}
    \mathbf{Z}_{\mathrm{opt}}
=\arg\max_{\mathbf{Z}\in\mathbb{C}^{N\times T}}
\Re\langle\mathbf{Z},\widehat{\mathbf{Y}}\rangle
~\text{s.t.}~
\sup_{\tau\in[0,1)}\bigl\|\mathbf{Z}^{\mathsf{H}}\mathbf{a}(\tau)\bigr\|_2\le1,
\end{align*}
and the support locations $\{\tau_k\}$ coincide with the frequencies at which the dual‐polynomial norm $\|\mathbf{Z}_{\mathrm{opt}}^{\mathsf{H}}\mathbf{a}(\tau)\|_2$ achieves its maximum of one. The corresponding snapshot amplitudes $\{\mathbf{b}_k\}$ are then obtained via least‐squares fitting of $\widehat{\mathbf{Y}}$ to the recovered spectral atoms.

\section{Numerical Simulations}\label{Sec:Simulation}

For our numerical simulations, we consider two choices of the lifting structure: (i) Hankel and (ii) double Hankel. These two choices result in two implementations of the proposed algorithm: WLi-EMaC for Hankle and WLi-DEMaC for double Hankle - see also \cite{BOKAEI2023109253}. The naming of these two algorithms is because of the similarity of the method with EMaC and DEMaC in \cite{chen2014robust,yang2021new}, respectively. This section presents several numerical experiments comparing WLi-EMaC and WLi-DEMaC with competing methods for the single snapshot DOA problem. More specifically, we consider (i) the EMaC method \cite{chen2014robust}, (ii) the double EMaC (DEMaC) method \cite{yang2021new}, (iii) the ANM technique \cite{bhaskar2013atomic},  and (iv) the grid-based implementation via the basis pursuit technique \cite{chen2001atomic}. Before proceeding further, let us clarify some general aspects of our simulations.

\noindent {\bf ULA setting:}
A requirement for all the above methods-- WLi-EMaC, WLi-DEMaC, EMaC, DEMaC, and ANM-- for applicability to the DOA estimation using a non-uniform grid is a completion procedure to form a uniform grid. After this completion step, the DOAs are obtained by applying the super-resolution technique in \cite{candes2014towards} over the estimated ULA. 
We assume the $\lambda/2$ spacing for this uniform grid, where $\lambda$ is the wavelength in \eqref{eq:atom}. Additionally, we assume the ULA consists of an odd number of array elements. These assumptions enable us to set the pencil parameter $d$ such that a square  Hankel matrix is achieved. 

\noindent {\bf Antenna/Source location:} 
SLA elements are selected uniformly at random from the ULA. For the source locations, two settings are considered:  (i) pre-determined source locations, where some sources are collocated, and (ii)  random source locations. For (i),  the pre-determined source angles and amplitudes are provided in Table~\ref{tab:source_setups}. For (ii), the DOA angles in \eqref{eq:atom} are  generated uniformly at random in the range $[-90^{\circ}, 90^{\circ}]$.  We note that this random selection ensures a general treatment of SLA configurations.

% ================================================
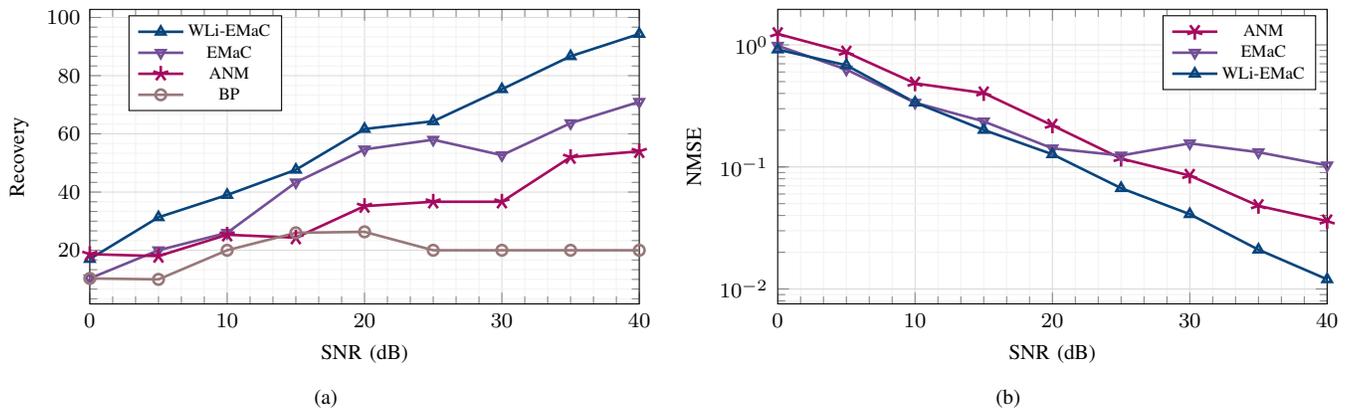
\begin{figure*}[!t]
\centering
\subfigure[]{\label{fig:snracc(a)}
    \begin{tikzpicture} 
    \begin{axis}[
        xlabel={SNR~(dB)},
        ylabel={Recovery},
        label style={font=\footnotesize},
        width=0.49\textwidth,
        height=5.5cm,
        xmin=0, xmax=40,
        ticklabel style = {font=\footnotesize},
        legend style={nodes={scale=0.65, transform shape}, at={(0.35,0.98)}}, 
        minor tick num=5,
        ymajorgrids=true,
        xmajorgrids=true,
        grid=both,
        grid style={line width=.1pt, draw=gray!10},
        major grid style={line width=.2pt,draw=gray!30},
    ]
    \addplot[
        color=darkcerulean,
        mark=triangle,
        line width=1pt,
        mark size=2pt,
        ]
    table[x=SNR,y=WEMaC]
    {data/Error_4.dat};
    \addplot[
        color=darklavender,
        mark=triangle,
        mark options = {rotate = 180},
        line width=1pt,
        mark size=2pt,
        ]
    table[x=SNR,y=EMaC]
    {data/Error_4.dat};
    \addplot[
        color=jazzberryjam,
        mark=star,
        line width=1pt,
        mark size=3pt,
        ]
    table[x=SNR,y=ANM]
    {data/Error_4.dat};
    \addplot[
        color=bazaar,
        mark=o,
        line width=1pt,
        mark size=2pt,
        ]
    ttable[x=SNR,y=BP]
    {data/Error_4.dat};
    \legend{WLi-EMaC, EMaC, ANM, BP};
    \end{axis}
    \end{tikzpicture}
    }\subfigure[]{\label{fig:snracc(b)}
	\begin{tikzpicture} 
    \begin{semilogyaxis}[
        xlabel={SNR~(dB)},
        ylabel={NMSE},
        label style={font=\footnotesize},
        ticklabel style = {font=\footnotesize},
        % legend cell align={left},
        width=0.49\textwidth,
        height=5.5cm,
        xmin=0, xmax=40,
        minor tick num=5,
        ymajorgrids=true,
        xmajorgrids=true,
        grid=both,
        grid style={line width=.1pt, draw=gray!15},
        major grid style={line width=.2pt, draw=gray!40},
        %ymin=1e-2, ymax=10,
        %xtick={0, 5, 10, 15, 20, 25, 30, 35, 40},
        %ytick={10, 1, 1e-1, 1e-2},
        legend style={nodes={scale=0.65, transform shape}, at={(0.98,0.98)}}, 
        %legend pos=east,
        %yminorgrids=true,
        %xmajorgrids=true,
        %grid style=dashed,
        grid=both,
        grid style={line width=.1pt, draw=gray!10},
        major grid style={line width=.2pt,draw=gray!30},
    ]
    \addplot[
        color=jazzberryjam,
        mark=star,
        mark options = {rotate = 180},
        line width=1pt,
        mark size=3pt,
        ]
    table[x=SNR,y=EMaC]
    {data/NMSE_4.dat};
    \addplot[
        color=darklavender,
        mark=triangle,
        mark options = {rotate = 180},
        line width=1pt,
        mark size=2pt,
        ]
    table[x=SNR,y=ANM]
    {data/NMSE_4.dat};
    \addplot[
        color=darkcerulean,
        mark=triangle,
        line width=1pt,
        mark size=2pt,
        ]
    table[x=SNR,y=WEMaC]
    {data/NMSE_4.dat};
    \legend{ANM, EMaC, WLi-EMaC};
    \end{semilogyaxis}
    \end{tikzpicture}
}
\caption{Scenario in Sec. \ref{sec:Robustness Against Noise}. (a) and (b) respectively represent the probability of correct DOA estimation (BP, Atomic, EMaC, and WLi-EMaC algorithms) and the normalized mean square error of the SLA interpolation to form the ULA (Atomic, EMaC, and WLi-EMaC algorithms) for a $20$-element SLA in terms of the input SNR. The location of $5$ sources is fixed such that two pairs are almost co-angled.
}
\label{fig:snracc}

\end{figure*}
% ================================================

\noindent  {\bf Estimation performance:} Generally, the array interpolation performance is evaluated through the normalized mean square (NMSE), which is defined as 
%---------------
\begin{align}
    \text{NMSE}:= \frac{\|\bm{y}-\hat{\bm{y}}\|_2^2}{\|\bm{y}\|_2^2},
\end{align}
%---------------
where $\bm{y}$ is the actual array samples and  $\hat{\bm{y}}$ is the interpolated array using Algorithm~\ref{alg:Low-rank-recovery}.  In Section~\ref{sec:The Effect of The Array Size} and \ref{sec:Robustness Against Noise}, a different metric is used.

\noindent
{\bf Grid-based DOA estimation:} In our simulations, we also compare the super-resolution methods with a grid-based DOA estimation. We uniformly divide the interval $[-1,1]$ into $2^{12}$ segments for the latter. 
The outcome of the angle estimation is, then, the angle corresponding to one such grid position.  
Note that the DOA problem in the grid-based setting simplifies to the standard compressed sensing problem. As such, the solution is determined through basis pursuit (BP). Therefore, the label BP in our figures in this section indicates this grid-based method.

Finally, for the parameters of the ADMM method, we set  $\rho = 10^{3}$ and $\gamma=10^5$. Note that the value of $\rho$ needs to be chosen sufficiently large to guarantee convergence; thus, larger values are also allowed. In contrast, $\gamma$ is tuned for each test.

\subsection{DOA Estimation}\label{sec:DOA Estimation}

In this experiment, we evaluate the performance for (i) four methods: BP, ANM, EMaC, and WLi-EMaC, and for  (ii)  two source structures under a noiseless setting ($\eta=0$). 
We consider an array aperture of $100\lambda/2$ and a $25$-element SLA. Source setups are chosen as the first two rows of Table~\ref{tab:source_setups}.
In Figure~\ref{fig:locations}, the estimated DOAs and the original DOAs are plotted. 
The results in Figures~\ref{fig:locations(a)}-\ref{fig:locations(d)} correspond to a setting with $8$ sources; two pairs of sources are intentionally placed very close.  
The WLi-EMaC has the best performance in this setting. 
First, BP could not resolve the seemingly collocated two sources. While ANM and EMaC were able to detect two separate sources in these locations, one of the estimated sources is greatly suppressed. Besides, both methods predict several small ghost sources.  Note that these ghost sources are primarily due to interpolation inaccuracies and model imperfections inherent in existing processes, which are further exacerbated by the sparse and non-uniform nature of the array; however, our proposed WLi-EMaC method significantly reduces their occurrence and amplitude by leveraging adaptive weighting to minimize interpolation errors.

% ================================================
%==========================
\begin{figure}[!t]
\centering
\subfigure[]{\label{fig:CVX_vs_ADMM(a)}
\begin{tikzpicture} 
\begin{semilogyaxis}[
xlabel={Number of samples},
ylabel={NMSE},
xmin=5, xmax=30,
label style={font=\footnotesize},
ticklabel style = {font=\footnotesize},
width=0.47\textwidth,
height=3.3 cm,
legend style={nodes={scale=0.6, transform shape}, at={(0.99,0.98)}}, 
ymajorgrids=true,
xmajorgrids=true,
]
\addplot[
color=jazzberryjam,
mark=triangle,
mark options = {rotate = 180},
line width=1pt,
mark size=2pt,
]
table[x=nSamp,y=cvx_err]
{data/cvx_vs_admm.dat};
\addplot[
color=darkcerulean,
mark=triangle,
line width=1pt,
mark size=2pt,
]
table[x=nSamp,y=ADMM_err]
{data/cvx_vs_admm.dat};
\legend{CVX , ADMM }
\end{semilogyaxis}
\end{tikzpicture}
}
\subfigure[]{\label{fig:CVX_vs_ADMM(b)}
\begin{tikzpicture} 
    \begin{semilogyaxis}[
    xlabel={SNR~(dB)},
    ylabel={NMSE},
    label style={font=\footnotesize},
    legend cell align={left},
    width=0.47\textwidth,
    height=3.3 cm,
    xmin=0, xmax=25,
    ticklabel style = {font=\footnotesize},
    legend style={nodes={scale=0.6, transform shape}, at={(0.99,0.98)}}, 
    ymajorgrids=true,
    xmajorgrids=true,
    ]
    \addplot[
    color=jazzberryjam,
    mark=triangle,
    mark options = {rotate = 180},
    line width=1pt,
    mark size=2pt,
    ]
    table[x=SNR,y=cvx_err]
    {data/cvx_vs_admm_noisy.dat};
    \addplot[
    color=darkcerulean,
    mark=triangle,
    line width=1pt,
    mark size=2pt,
    ]
    table[x=SNR,y=ADMM_err]
   {data/cvx_vs_admm_noisy.dat};
  \legend{CVX , ADMM }
  \end{semilogyaxis}
\end{tikzpicture}
}
\caption{Performance of the CVX and ADMM implementation in   Sec. \ref{ADMM Algorithm versus Convex Implementation}.
of WLi-EMaC in array interpolation with aperture size $48\lambda/2$ for two scenarios. In Figure \ref{fig:CVX_vs_ADMM(a)}, two algorithm performances are compared in noiseless for SLA with $5$ to $30$ number of elements in terms of NMSE. In Figure~\ref{fig:CVX_vs_ADMM(b)}, the performance of the algorithms is compared for different SNRs in terms of NMSE. In both cases, $4$ sources are generated with random directions.}
\label{fig:CVX_vs_ADMM}
\end{figure}
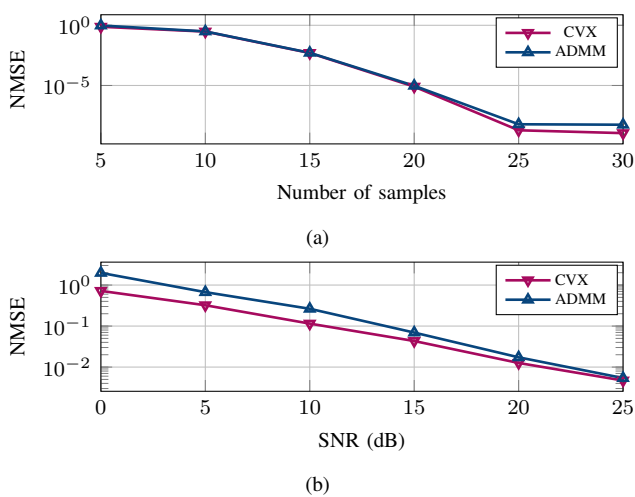
%==========================

%==========================
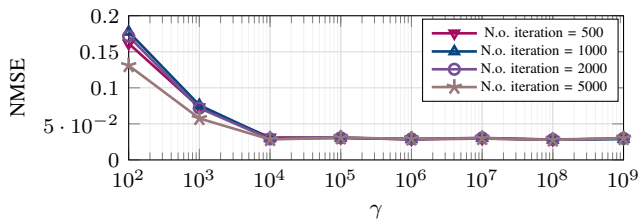
\begin{figure}[t!]
\centering
\begin{tikzpicture} 
\begin{semilogxaxis}[
xlabel={$\gamma$},
ylabel={NMSE},
label style={font=\footnotesize},
width=0.45\textwidth,
ticklabel style = {font=\footnotesize},
height=3.5cm,
xmin=1e2, xmax=1e9,
ymin=0, ymax=0.2,
xtick={1e2,1e3,1e4,1e5,1e6,1e7,1e8,1e9},
ytick={0, 0.05, 0.1, 0.15, 0.2},
legend style={nodes={scale=0.55, transform shape}, at={(0.99,0.98)}}, 
grid=both,
grid style={line width=.1pt, draw=gray!10},
major grid style={line width=.2pt,draw=gray!30},
]
\addplot[
color=jazzberryjam,
mark=triangle,
mark options = {rotate = 180},
line width=1pt,
mark size=2pt,
]
table[x=gamma,y=itr_500]
{data/gamma_effect.dat};
\addplot[
color=darkcerulean,
mark=triangle,
line width=1pt,
mark size=2pt,
]
table[x=gamma,y=itr_1000]
{data/gamma_effect.dat};
\addplot[
color=darklavender,
mark=o,
line width=1pt,
mark size=2pt,
]
table[x=gamma,y=itr_2000]
{data/gamma_effect.dat};
\addplot[
color=bazaar,
mark=star,
line width=1pt,
mark size=3pt,
]
table[x=gamma,y=itr_5000]
{data/gamma_effect.dat};
\legend{N.o. iteration = 500, N.o.  iteration = 1000, N.o.  iteration = 2000, N.o. iteration = 5000};
\end{semilogxaxis}
\end{tikzpicture}
\caption{ Performance of ADMM algorithm  in Sec. \ref{ADMM Algorithm versus Convex Implementation}in solving WLi-EMaC for different $\gamma$s for following scenario: $30$ sample SLA with $SNR = 20 {\rm dB}$ is chosen randomly from ULA with $60\lambda/2$ aperture size. Source setup is the last row of Tabel \ref{tab:source_setups}, and $\rho$ is $10^3$ in all simulations}
\label{fig:gamma_effect}
\end{figure}
% ================================================

In Figures~\ref{fig:locations(e)}-\ref{fig:locations(h)}, we have included $9$ sources with two almost collocated triplets.  
While WLi-EMaC detects all source locations correctly, with unbalanced amplitudes, it still performs best among the considered methods.

\subsection{The Effect of the Array Size}\label{sec:The Effect of The Array Size}

To investigate the effect of the number of array elements in the overall performance, again, we consider a uniform array with aperture  $58\lambda/2$. We further form the SLA with $5$ to $26$ elements. We consider scenario (c) in Table~\ref {tab:source_setups}.

The first, second, and third sources are seemingly 
collocated, as well as four and five.  Note that, for all $\theta_i$ and $\theta_j$, $|\sin(\theta_i) - \sin(\theta_j)|\geq 0.01$. For this scenario, DOA estimation is considered successful if there is only one estimated angle $\hat{\theta}$ for the angle $\theta$ that satisfies $|\sin(\theta) - \sin(\hat{\theta})| \leq 0.005$.

In Figure~\ref{fig:accuracy(a)}, the percentage of correct source recovery concerning the size of the SLA is plotted. The performance is averaged over $100$ random realizations of the SLA elements for each curve. As expected, the recovery percentage increases as the elements increase. The results show that the EMaC-based algorithms outperform the other two methods. 
In addition, the weighted versions of EMaC and DEMaC -- WLi-EMaC and WLi-DEMaC --  have slightly better performances. 

In Figure~\ref{fig:accuracy(b)}, the NMSE of the estimated unobserved ULA elements in the SLA is depicted. These curves represent the performance of the involved matrix completion procedure (therefore, the BP method is excluded here). Weighted-based methods have smaller NMSE values than their corresponding structures in most parts. Also, DEMaC performs better than EMaC for most sample sizes; however, WLi-EMaC outperforms DEMaC for larger sample sizes.

\subsection{Robustness Against Noise}\label{sec:Robustness Against Noise}

To investigate the methods' estimation performance in a noisy setting, we consider the ULA with $100\lambda/2$ aperture and fix the size of the SLA as $20$. We consider the $5$-source setup of the previous subsection and use the same criteria for successful DOA estimation. In this section, we investigate WLi-EMaC performance. For each SNR value, we consider $100$ random realizations of the SLA with $20$ elements.

In Figures~\ref{fig:snracc(a)} and \ref{fig:snracc(b)}, the percentage of true source recovery and the average NMSE in the matrix completion procedure are shown, respectively. While the performance of all the methods degrades gradually as the SNR decreases, WLi-EMaC has considerably superior performance.

Note that the proposed WLi-EMaC and WLi-DEMaC are evaluated in scenarios with varying numbers of sources,e.g., Figures~\ref{fig:locations},\ref{fig:accuracy}, and \ref{fig:snracc}. The array completion method consistently demonstrated accurate DOA recovery without requiring prior knowledge of the number of sources, highlighting its adaptability and flexibility in practical applications.

\begin{figure}[!t]
\centering
	\begin{tikzpicture} 
    \begin{semilogyaxis}[
        xlabel={Number of snapshots $T$ },
        ylabel={NMSE},
        label style={font=\footnotesize},
        ticklabel style = {font=\footnotesize},
        % legend cell align={left},
        width=0.49\textwidth,
        height=5.5cm,
        xmin=1, xmax=10,
        ymin=1e-1, ymax=1,
        minor tick num=5,
        ymajorgrids=true,
        xmajorgrids=true,
        grid=both,
        grid style={line width=.1pt, draw=gray!15},
        major grid style={line width=.2pt, draw=gray!40},
        legend style={nodes={scale=0.65, transform shape}, at={(0.98,0.98)}}, 
        %legend pos=east,
        %yminorgrids=true,
        %xmajorgrids=true,
        %grid style=dashed,
        grid=both,
        grid style={line width=.1pt, draw=gray!10},
        major grid style={line width=.2pt,draw=gray!30},
    ]
    \addplot[
        color=jazzberryjam,
        mark=star,
        mark options = {rotate = 180},
        line width=1pt,
        mark size=3pt,
        ]
    table[x=SNAP,y=EMaC]
    {data/NMSE_6.dat};
    \addplot[
        color=darklavender,
        mark=triangle,
        mark options = {rotate = 180},
        line width=1pt,
        mark size=2pt,
        ]
    table[x=SNAP,y=ANM]
    {data/NMSE_6.dat};
    \addplot[
        color=darkcerulean,
        mark=triangle,
        line width=1pt,
        mark size=2pt,
        ]
    table[x=SNAP,y=WEMaC]
    {data/NMSE_6.dat};
    \legend{ANM, EMaC, WLi-EMaC};
    \end{semilogyaxis}
    \end{tikzpicture}

\caption{Figure represents the normalized mean square error of the SLA interpolation to form the ULA (Atomic, EMaC, and WLi-EMaC algorithms) for a $20$-element SLA in terms of the number of snapshots $T$. To assess the algorithm's performance for co-angled sources, the location of two pairs of $5$ sources is fixed so they become almost co-angled.
}
\label{fig:snapacc}

\end{figure}
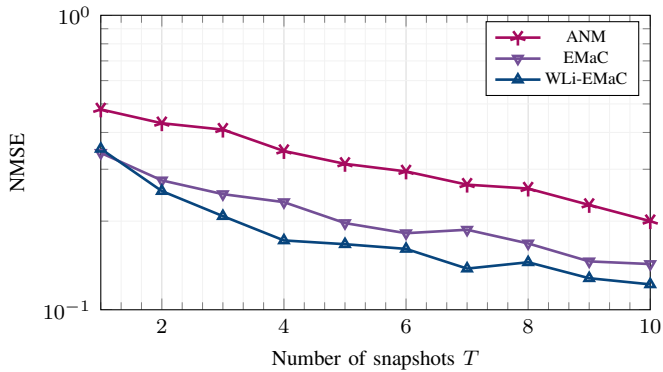

Moreover, to assess the impact of the number of snapshots $T$ on DOA estimation under a fixed noise level, we fix \(\mathrm{SNR}=10\)\,dB and vary \(T\) from $1$ to $10$. For each \(T\), we generate $100$ independent SLA realizations (each with 20 sensors randomly selected from the \(100\lambda/2\) ULA) and simulate the same 5-source scenario. Figure~\ref{fig:snapacc} shows the average NMSE of the matrix‐completion step as a function of \(T\).

As \(T\) increases, with a single snapshot the NMSE remains high, whereas by \(T=10\) snapshots the NMSE falls around $-10$~dB. These results confirm that WLi-EMaC effectively leverages temporal diversity to enhance robustness against noise, achieving near-optimal performance with a moderate number of snapshots.

% ================================================
% ================================================

\subsection{ADMM Algorithm Versus Convex Implementation}\label{ADMM Algorithm versus Convex Implementation}

In closing, we compare the performance of the ADMM algorithm versus the CVX toolbox~\cite{cvx} in terms of accuracy and complexity for the noiseless and noisy scenarios. In the first setup, we consider a ULA with aperture  $48\lambda/2$, and form an SLA  with $5$ to $30$ elements. In the noisy case, we create a $25$-element SLA and SNRs between $0$ dB and $25$ dB. Furthermore, the direction of $4$ sources is chosen uniformly at random in both scenarios.   In Figure~\ref{fig:CVX_vs_ADMM}, the performance of the ADMM algorithm compared to the  CVX Toolbox for both noiseless and noisy cases is shown.

We observe that ADMM is slightly less accurate with the considered number of iterations; however, the computational cost is drastically lower.  More specifically, ADMM was around $34$ times faster than the CVX Toolbox in this experiment.  To investigate the effect of $\gamma$ in the convergence of the ADMM algorithm, we consider a ULA with $60 \lambda/2$ aperture size from a $30$-element SLA and add noise with SNR $= 20$ dB to the SLA samples.  We consider the $3$-source setup in the last row of Table~\ref{tab:source_setups}, set $\rho = 10^3$, and sketch the normalized mean-squared error of the recovered ULA by the ADMM algorithm for $500$, $1000$, $2000$, and $5000$ iterations in Figure~\ref{fig:gamma_effect}.  Each curve is averaged over $100$ Monte Carlo trials.  In Figure~\ref{fig:gamma_effect}, we observe that for $\gamma \geq 10^4$, the ADMM algorithm with the considered number of iterations reaches its converging point.

\subsection{Performance Comparison with  Analytical Lower Bounds}\label{Sec:bounds}

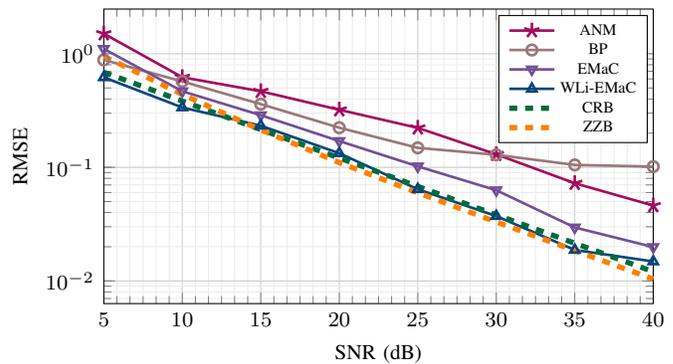
\begin{figure}[t]
\centering
\begin{tikzpicture} 
\begin{semilogyaxis}[
xlabel={SNR (dB)},
ylabel={RMSE},
xmin=5, xmax=40,
label style={font=\footnotesize},
ticklabel style = {font=\footnotesize},
width=0.49\textwidth,
height=5.5 cm,
minor tick num=5,
legend style={nodes={scale=0.6, transform shape}, at={(0.99,0.98)}}, 
ymajorgrids=true,
xmajorgrids=true,
grid=both,
grid style={line width=.1pt, draw=gray!15},
major grid style={line width=.2pt, draw=gray!40},
]
\addplot[
color=jazzberryjam,
mark=star,
line width=1pt,
mark size=3pt,
]
table[x=SNR,y=anm]
{data/bound.dat};
\addplot[
color=bazaar,
mark=o,
line width=1pt,
mark size=2pt,
]
table[x=SNR,y=bp]
{data/bound.dat};
\addplot[
color=darklavender,
mark=triangle,
mark options = {rotate = 180},
line width=1pt,
mark size=2pt,
]
table[x=SNR,y=Emac]
{data/bound.dat};
\addplot[
color=darkcerulean,
mark=triangle,
line width=1pt,
mark size=2pt,
]
table[x=SNR,y=Wli]
{data/bound.dat};
\addplot[
color=cadmiumgreen,
dashed,
line width=2pt,
mark size=3pt,
]
table[x=SNR,y=crbd]
{data/bound.dat};
\addplot[
color=orange,
dashed,
line width=2pt,
mark size=3pt,
]
table[x=SNR,y=zzb]
{data/bound.dat};
\legend{ANM, BP, EMaC, WLi-EMaC, CRB, ZZB }
\end{semilogyaxis}
\end{tikzpicture}
\caption{ %{\color{blue!70!black!80} 
Comparison of the RMSE performance of the proposed methods (WLi-EMaC and WLi-DEMaC) with existing state-of-the-art DOA estimation algorithms, plotted against the Cramér-Rao Bound (CRB) and Ziv-Zakai Bound (ZZB). The scenario involves two closely spaced sources with angular separation \( |\sin(\theta_i) - \sin(\theta_j)| = 0.015 \), using an array with an aperture of \( 58\lambda/2 \) comprising $25$ elements.}
\label{fig:bounds}
\end{figure}

For the last simulation, we evaluate the performance of the proposed DOA estimation method compared to state-of-the-art methods. Here,  to compute the DOA estimation performance, we use root mean square error (RMSE) as a metric, which is defined by 
%------------
\begin{align}
    \text{RMSE} := \sqrt{\frac{1}{K}\sum_{k=1}^K \mathbb{E}\{(\hat{\theta}_k-\theta_k)^2\}}, 
\end{align}
%------------
where $\theta_k$ is the actual angle of source $k$ and $\hat{\theta}_k$ is the estimation thereof.  We also provide RMSE of the methods against the Cramér-Rao bound (CRB) \cite{cramer1999mathematical} and the Ziv-Zakai bound (ZZB) \cite{zhang2022ziv} to assess theoretical and practical limitations. The scenario considers two closely spaced sources with a separation of \( |\sin(\theta_i) - \sin(\theta_j)| = 0.015 \) and an array with an aperture size of \( 58\lambda / 2 \) consisting of $25$ elements. Each experiment has been performed over $500$ Monte Carlo trials. This setup provides insight into different algorithms' resolution capabilities and robustness under challenging conditions. The result is depicted in Figure~\ref{fig:bounds}. We observe that the proposed WLi-EMaC and WLi-DEMaC outperform the other methods in terms of RMSE. However, in the low SNR regime ($<10$ dB), the weighted schemes appear to exceed the theoretical lower bounds; we speculate this artifact is due to the finite number of Monte Carlo trials.

% ================================================
%==========================
%==========================
% ================================================

% ======================================
\section{Conclusion}\label{Sec:Conclude}
% ======================================

In this paper,  we proposed a method for the direction of arrival (DOA) estimation for non-uniformly spaced linear antenna arrays. Our method comprised three steps: (i) the array samples are lifted to a chosen structured matrix, such as Hankel or Toeplitz. Then, (ii) left and right weighting matrices were determined to reflect the sample informativeness, and (iii) the weighted and lifted structure was used to estimate the noiseless uniformly-spaced array samples through low-rank matrix completion. For a given choice of the lifting structured matrix, this weighting method generalizes other low-rank matrix completion methods introduced in the literature, such as EMaC  -- for Hankel lifting-- and DEMaC -- for double Hankel. Numerical results showed that the weighted lifted (WLi-) method generally outperforms the case without weighting. In other words, WLi-EMaC/WLi-DEMac outperforms EMaC and DEMaC regarding the NMSE metric.

\appendix

\subsection{Proof of Lemma~\ref{prop:Lev_score}}\label{Lem:OptmialWeight}

Let us begin by describing the proof strategy from a high-level perspective. 
To obtain an upper-bound for the leverage scores, we first use a surrogate function for simplifying the minimization in \eqref{eq:weightedUpper}. For this surrogate function,  the upper bound allows us to obtain a closed-form solution for the weight matrices. Accordingly, we prove the upper-bound for the leverage scores using the resultant weight matrices; therefore, the upper bound is valid for optimization in \eqref{eq:weightedUpper} as we used a surrogate approximation.

Towards obtaining a closed-form solution, instead of solving  \eqref{eq:weightedUpper}, one can minimize an upper-bound of the objective function as 
% -------------
\begin{align}
\label{eq:weightsum}
\underset{\substack{  \mathbf{W}_L, \mathbf{W}_R \in \mathbb{R}_{+}} }{\rm minimize} 
 \sum_{n \not\in \Omega}{  \|\mathbf{W}_L\mathbf{A}_n \|_{\rm F}^2 } +  {  \|\mathbf{W}_R\mathbf{A}_n^{\mathsf{T}} \|_{\rm F}^2 } \\ 
~~ {\rm s.t.}~~
\|\mathbf{W}_L   \|_{\rm F}^2 = 1,  \|\mathbf{W}_R\|_{\rm F}^2 = 1.
\nonumber
\end{align}
% -------------
or equivalently
% -------------
\begin{align}
\nonumber
\underset{\substack{ \mathbf{W}_L, \mathbf{W}_R \in \mathbb{R}_{+}} }{\rm minimize} 
 {\rm tr}\big(\mathbf{W}_L^{\mathsf{T}}\mathbf{W}_L\mathbf{G}_{\Omega}^L\big)  +    {\rm tr}\big(\mathbf{W}_R^{\mathsf{T}}\mathbf{W}_R\mathbf{G}_{\Omega}^R\big), \\
~~{\rm s.t.}~~
\|\mathbf{W}_L\|_{\rm F}^2 = 1,  \|\mathbf{W}_R\|_{\rm F}^2 = 1,  \nonumber
\end{align}
% -------------
where  
\begin{align}
\nonumber
\mathbf{G}_{\Omega}^L  := \sum_{n \not\in \Omega} \mathbf{A}_n\mathbf{A}_n^{\mathsf{T}}, \quad
\mathbf{G}_{\Omega}^R  := \sum_{n \not\in \Omega} \mathbf{A}_n^{\mathsf{T}}\mathbf{A}_n.  
\end{align}
The latter minimization problem consists of two separate objective functions, and it can be divided into two separate optimization problems as below:
% -------------
\begin{subequations}
\label{eq:weightedUpperLR}
\begin{align}
\label{eq:weightedUpperL}
 \mathbf{w}_L^{*} &:= \underset{\substack{ \mathbf{w}_L\in \mathbb{R}_{+}} }{\rm argmin} ~
{\rm tr} \Big( \mathbf{w}_L^{\mathsf{T}} {\rm diag} (\mathbf{G}_{\Omega}^L)  \Big),  \nonumber \\
&  \quad \quad ~~{\rm s.t.}~~
\sum_{i \in [d]}w_{L,i} = 1, \\
\label{eq:weightedUpperR}
\mathbf{w}_R^{*} &:= \underset{\substack{ \mathbf{w}_R\in \mathbb{R}_{+}}}{\rm argmin}~{\rm tr}\Big( \mathbf{w}_R^{\mathsf{T}} {\rm diag} (\mathbf{G}_{\Omega}^R) \Big),\nonumber \\
& \quad \quad ~~{\rm s.t.}~~
\sum_{i \in [N-d+1]}w_{R,i} = 1, 
\end{align}
\end{subequations}
% -------------
where 
%--------------
$\mathbf{w}_L  = [w_{L,1},\ldots, w_{L,d}]^{\mathsf{T}}$,$
\mathbf{w}_R = [w_{R,1},\ldots, w_{R,N-d+1}]^{\mathsf{T}}$.
%--------------
The minimum to \eqref{eq:weightedUpperLR} occurs when all elements of $\mathbf{w}_L$ and $\mathbf{w}_R$ become zero except for the ones selecting the minimum diagonal value of $\mathbf{G}_{\Omega}^L$ and $\mathbf{G}_{\Omega}^R$, respectively. The minimum diagonal value of $\mathbf{G}_{\Omega}^L$ and $\mathbf{G}_{\Omega}^R$ can be bounded by the maximum diagonal value, which can be written as
% -------------
\begin{subequations}
\begin{align}
0 \leq \min\{{\rm diag} (\mathbf{G}_{\Omega}^L) \} &  \leq 2\log(d), \\
0 \leq  \min\{{\rm diag} (\mathbf{G}_{\Omega}^R) \} &  \leq 2\log(d).
\end{align}
\end{subequations}
% -------------
Note that given $\Omega$, the minimum values can even become zero. Then, by substituting $\mathbf{w}_L^*$ and $\mathbf{w}_R^*$ into \eqref{eq:mutildDef} and \eqref{eq:weightsum} and summing up over $n$, we obtain 
% -------------
\begin{align}
    \nonumber
    \frac{\sum_{n \in [N]}\tilde{\mu}_n\tilde{K}}{N} &  \leq \sum_{n \in [N]} \|\mathbf{W}_L^{*}\mathbf{A}_n\|_{\rm F}^2 +  \|\mathbf{W}_R^{*}\mathbf{A}_n^{\mathsf{T}}\|_{\rm F}^2, \\
     & \leq  4\log(N), \label{eq:lowermu}
\end{align}
% -------------
for $b = 4$; this concludes the proof. 

\bibliographystyle{IEEEtran}
\bibliography{IEEEabrv,Ref}

\end{document}